\documentclass[twocolumn]{aastex62}
\usepackage{amsmath,amstext}
\usepackage[T1]{fontenc}
\usepackage{natbib}
\usepackage{sidecap}

\begin{document}

\title{Dynamical instabilities in systems of multiple short-period planets are likely driven by secular chaos: a case study of Kepler-102} 
\shorttitle{secular chaos in multi-planet systems}

\author[0000-0001-8736-236X]{Kathryn Volk}
\correspondingauthor{Kathryn Volk}
\email{kvolk@lpl.arizona.edu}
\affiliation{Lunar and Planetary Laboratory, The University of Arizona, 1629 E University Blvd, Tucson, AZ 85721}

\author[0000-0002-1226-3305]{Renu Malhotra}
\affil{Lunar and Planetary Laboratory, The University of Arizona, 1629 E University Blvd, Tucson, AZ 85721}

%%%%%%%%%%%%%%%%%%%%%%%%%%%%%%%%%%%%%%%%%%%
\begin{abstract}
We investigated the dynamical stability of high-multiplicity {\it Kepler} and {\it K2} planetary systems. Our numerical simulations find instabilities in $\sim20\%$ of the cases on a wide range of timescales (up to $5\times10^9$ orbits) and over an unexpectedly wide range of initial dynamical spacings.
To identify the triggers of long-term instability in multi-planet systems, we investigated in detail the five-planet Kepler-102 system.
Despite having several near-resonant period ratios, we find that mean motion resonances are unlikely to directly cause instability for plausible planet masses in this system.
Instead, we find strong evidence that slow inward transfer of angular momentum deficit (AMD) via secular chaos excites the eccentricity of the innermost planet, Kepler-102 b, eventually leading to planet-planet collisions in $\sim80\%$ of Kepler-102 simulations.
Kepler-102 b likely needs a mass $\gtrsim0.1M_{\earth}$, hence a bulk density exceeding about half Earth's, in order to avoid dynamical instability.
To investigate the role of secular chaos in our wider set of simulations, we characterize each planetary system's AMD evolution with a ``spectral fraction" calculated from the power spectrum of short integrations ($\sim5\times10^6$ orbits). 
We find that small spectral fractions ($\lesssim0.01$) are strongly associated with dynamical stability on long timescales ($5\times10^9$ orbits) and that the median time to instability decreases with increasing spectral fraction.
Our results support the hypothesis that secular chaos is the driver of instabilities in many non-resonant multi-planet systems, and also demonstrate that the spectral analysis method is an efficient numerical tool to diagnose long term (in)stability of multi-planet systems from short simulations.

\end{abstract}
%%%%%%%%%%%%%%%%%%%%%%%%%%%%%%%%%%%%%%%%%%%


%%%%%%%%%%%%%%%%%%%%%%%%%%%%%%%%%%%%%%%%%%%
%%%%%%%%%%%%%%%%%%%%%%%%%%%%%%%%%%%%%%%%%%%
\section{Introduction} \label{sec:intro}
%%%%%%%%%%%%%%%%%%%%%%%%%%%%%%%%%%%%%%%%%%%
%%%%%%%%%%%%%%%%%%%%%%%%%%%%%%%%%%%%%%%%%%%
The {\it Kepler} space telescope carried out a large systematic exoplanet survey during the {\it Kepler}~\citep{Borucki:2010} and {\it K2} missions~\citep{Howell:2014}, and has provided a wealth of data on planets and planetary systems in the Galaxy.  
A large subset, about  40\%, of the {\it Kepler} planet candidates are in systems with two or more planet candidates \citep[e.g.,][]{Coughlin:2016}.
Studies of the {\it Kepler} data on multiple-planet systems have concluded that planetary systems are typically coplanar to within a few degrees~\citep[e.g.,][]{Lissauer:2011, Tremaine:2012, Fang:2012, Johansen:2012, Figueira:2012, Fabrycky:2014} and that they generally have low eccentricities as well \citep[e.g.,][]{Xie:2016,VanEylen:2015}. 
In contrast with the Solar system, most of the {\it Kepler} systems are tightly packed, within less than 0.5 au of their stellar host \citep[e.g.,][]{Lissauer:2011,Fressin:2013}. 
Theoretical studies of dynamical stability of systems with similar masses and orbital spacings to those of the observed Kepler systems conclude that such systems are close to the threshold of instability~\citep[e.g.,][]{Fang:2013}. 
\citet{Pu:2015} and \citet{Volk:2015} have suggested this could be because inner planet systems form with even tighter spacings and higher multiplicity, then undergo a sequence of dynamical instabilities that pare down the numbers of planets and widen their orbital spacings, a process involving  planet-planet collisions and mergers.

\begin{figure}[ht]
\centering
\includegraphics[width=3in]{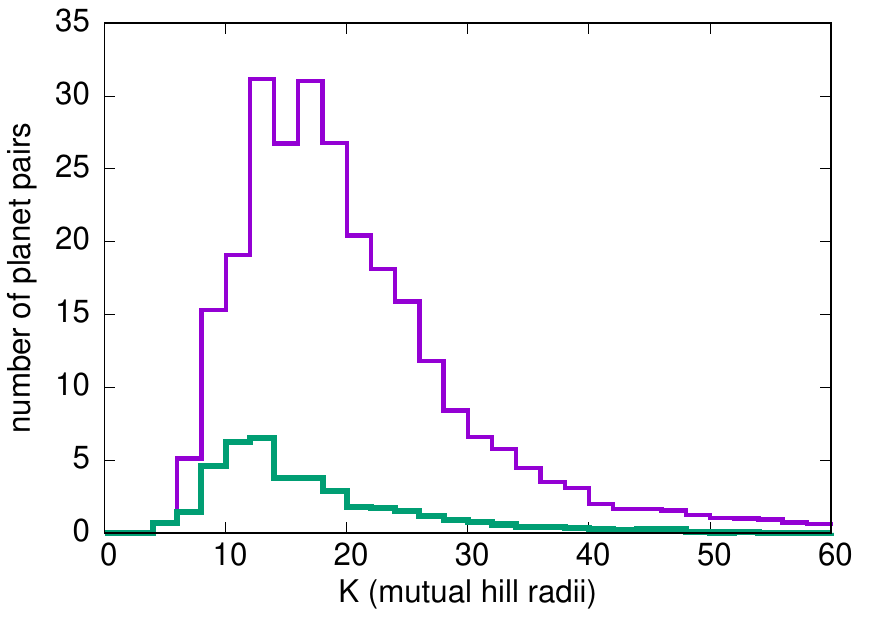}
\caption{Distribution of estimated  dynamical spacing, $K$, for all pairs of adjacent {\it Kepler} and {\it K2} confirmed planet candidates in systems with four or more planets (purple histogram). The masses of the planets are taken from the \cite{Wolfgang:2016} statistical mass radius relationship; each system is sampled 100 times to generate the histogram, but the total weight of each adjacent planet pair in the above histogram is set to one. The green histogram shows the subset of simulated planet pairs in these systems that collide due to instabilities.
\label{f:delta}}
\end{figure}

The orbital spacings in planetary systems are usefully described in units of the mutual Hill radius ($R_{mH}$) of adjacent planets. 
For two planets of masses $m_1$ and $m_2$ orbiting a star of mass $M_*$ in orbits of semimajor axes $a_1$ and $a_2$, the parameter $K$ describes the number of mutual Hill radii separating the planetary orbits:
%%%%%%%%%%%%%%%%%%%%%%%%%%%%%%%%%%%%%%%%%%%
\begin{equation}
K = \frac{|a_1 - a_2|}{R_{mH}} = \frac{2|a_1 - a_2|}{a_1+a_2}\left[\frac{3M_*}{m_1+m_2}\right]^{1/3}.
\label{e:K}\end{equation}
%%%%%%%%%%%%%%%%%%%%%%%%%%%%%%%%%%%%%%%%%%%
 The dynamical spacing requirement for stable co-planar, low-eccentricity two-planet systems is relatively simple and closely follows that for the coplanar restricted three body problem \citep[see, e.g.,][and references therein]{Gladman:1993}, 
\begin{equation}
K > K_{min} = 2\sqrt3.
\label{e:Kmin}
\end{equation}
In contrast with two planet systems, the stability in higher-multiplicity systems is poorly understood and is generally expected to be of a statistical nature, although we can expect that the minimum dynamical separation of Eq.~\ref{e:Kmin} must be satisfied~\citep{Malhotra:2015}.
Empirical scaling laws for the relationship between stability times and dynamical separations have been derived with numerical integrations of evenly-spaced, equal-mass planets \citep[e.g.,][]{Chambers:1996,Smith:2009,Funk:2010,Obertas:2017} as well as for inhomogeneous systems (that is, those with a dispersion in planet masses, orbital spacings and orbital eccentricities and inclinations, \cite{Pu:2015}).
However, these empirical scaling relationships have limited applicability to the observed multi-planet distribution because they mostly apply to instabilities at dynamical separations of $K\lesssim$~10--15, whereas the observed multi-planet systems have a much broader distribution of $K$ (see Figure~\ref{f:delta}). 
Additionally, although empirical rules for dynamical (in)stability have been found in numerical simulations, there is relatively little detailed understanding of the underlying mechanisms that generate the instabilities in simulated planetary systems.

%%%%%%%%%%%%%%%%%%%%%%%%%%%%%%%%%%%%%%%%%%%
%%%%%%%%%%%%%%%%%%%%%%%%%%%%%%%%%%%%%%%%%%%

In the present work, we expand upon these previous studies by using the observed {\it Kepler} and {\it K2} planetary architectures as a starting point for studying the underlying source of dynamical stability and instability in the diverse set of observed compact systems of small planets. 
Figure~\ref{f:delta} shows the estimated $K$ distribution of all the currently catalogued adjacent pairs of planets observed during the {\it Kepler} and {\it K2} missions in systems with 4 or more planets (population statistics taken from the NASA Exoplanet Archive\footnote{\url{exoplanetarchive.ipac.caltech.edu}}). 
 The observational data can be used to measure the fractional orbital separations from the orbital periods, but do not directly provide planet masses.
Therefore, to compute the $K$ values, we use the \cite{Wolfgang:2016} statistical mass-radius relationship to assign a range of possible masses to each observed planet based on the planet radius reported in the NASA Exoplanet Archive's Composite Planet Data table. 
For each observed planetary system, we generate 100 versions of that system from the mass-radius relationship and calculate $K$ for each adjacent pair in the system. 
The resulting histogram of $K$ values (in which each observed planet pair has a weight of one) is shown in Figure~\ref{f:delta}.

Considering that most of the {\it Kepler} multi-planet host stars are several gigayears old \citep[e.g.,][]{SilvaAguirre:2015}, it is notable that so many planet pairs appear to have quite small values of $K$, $K\lesssim15$, where the above-mentioned scaling laws would suggest dynamical lifetimes shorter than gigayears.
 Perhaps some of the planets at small spacings are spurious detections or have masses smaller than we have assumed.
However, we find that even large dynamical spacings do not guarantee stability.
In our numerical simulations of these systems (described in Section~\ref{s:keplermultis}), $\sim20\%$ of simulated systems became unstable on timescales of a few billion orbits ($\lesssim100$~Myr).
Approximately half of the instabilities occur between planet pairs that begin with $K>15$ (see the green histogram in Figure~\ref{f:delta}), separations predicted to be stable based on the simple scaling laws mentioned above.

This motivates us to investigate the dynamics of these planetary systems in order to identify the source of dynamical instabilities as well as markers for dynamically stable architectures.
The rest of this paper is organized as follows. 
In Section~\ref{s:keplermultis}, we describe our simulation results for the distribution of dynamical lifetimes of multi-planet systems spanning the range of orbital architectures similar to those of the observed sample of {\it Kepler} multis. 
In Section~\ref{s:casestudy}, we describe our case study of the Kepler-102 system: we show with analytical and numerical estimates that the overlap of mean motion resonances is unlikely to be the direct cause of dynamical instabilities in this system, we investigate the secular dynamics in this system, and we use an analysis of the system's angular momentum deficit (AMD) to show that secular chaos is the likely cause of dynamical instabilities in simulations of this system. 
In Section~\ref{s:spectra} we describe the usefulness of spectral analysis of relatively short simulations to diagnose long-term stable and unstable planetary architectures; based on this analysis, we tentatively conclude that secular chaos likely drives instability in simulations of many of the observed Kepler and K2 multiplanet systems. 
We summarize in Section~\ref{s:summary}.

%%%%%%%%%%%%%%%%%%%%%%%%%%%%%%%%%%%%%%%%%%%
%%%%%%%%%%%%%%%%%%%%%%%%%%%%%%%%%%%%%%%%%%%
\section{Dynamical lifetimes of Kepler multiplanet systems}\label{s:keplermultis}
%%%%%%%%%%%%%%%%%%%%%%%%%%%%%%%%%%%%%%%%%%%
%%%%%%%%%%%%%%%%%%%%%%%%%%%%%%%%%%%%%%%%%%%

To investigate the potential drivers of instability in realistic multiplanet systems, we performed a large suite of simulations based on the observed properties of the {\it Kepler} and {\it K2} systems with four or more confirmed planets and an estimated stellar mass.
A complete analysis of these simulations is deferred to a future paper, but here we use some of their basic results to motivate a case study (Section~\ref{ss:K102}) of the detailed dynamics that drive instabilities in multi-planet systems.

For each observed planetary system in our sample, we use the properties of the system listed in the NASA Exoplanet Archive's Composite Planet Data table combined with a statistical mass-radius relationship as the basis for our simulation initial conditions. 
From this database, we take the stellar mass, planetary radius, and orbital periods (along with all associated uncertainties) for each system. 
The planetary radii and uncertainties are fed into the \cite{Wolfgang:2016} statistical mass-radius relationship\footnote{Their code is available at \url{github.com/dawolfgang/MRrelation}} to generate 100 masses for each planet; this includes 50 masses calculated using \cite{Wolfgang:2016}'s transit timing variation priors and 50 calculated using the radial velocity priors. 
For each generated set of planetary masses, a stellar mass is selected uniformly from within the uncertainties (to ensure we fully sample the most likely stellar mass range with our relatively small set of simulations), and the observed planetary orbital periods are then used to calculate a semimajor axis for each planet. 
The eccentricities and inclinations of each planet are then selected from rayleigh distributions of widths $\sigma_e=0.02$ and $\sigma_i=1.4^\circ$, respectively; these widths are consistent with the estimated intrinsic eccentricity and inclination widths for {\it Kepler} multiplanet systems \citep[e.g.,][]{Hadden:2014, Xie:2016,VanEylen:2015}. 
The orbital angles (argument of perihelion, longitude of ascending node, and mean anomaly) for each planet are randomized in the range 0 to $2\pi$. 
These initial conditions do not represent the exact Kepler systems, but represent plausible variations of each observed planetary system.

The resulting $\sim6000$ sets of planetary system initial conditions were then integrated for $5\times10^9$ orbits of the innermost planet (corresponding to $\sim10-100$~Myr timescales, depending on the system). 
The integrations were performed using the {\sc{mercurius}} integrator (based on \citealt{Chambers:1999}) within {\sc{rebound}} \citep{Rein:2012}. A post-Newtonian general relativity correction was added to the code \citep[using the same prescription as in][]{Lissauer:2011}. 
The initial timestep was set to 1/40th of the innermost planet's orbital period, and the integrator switches to an adaptive timestep routine to handle close encounters between planets. 
The integrations were stopped if two planets collided (the collision radius was set to the physical radius of the planets).

%%%%%%%%%%%%%%%%%%%%%%%%%%%%%%%%%%%%%%%%%%%
\begin{figure}
    \centering
    \includegraphics[width=3in]{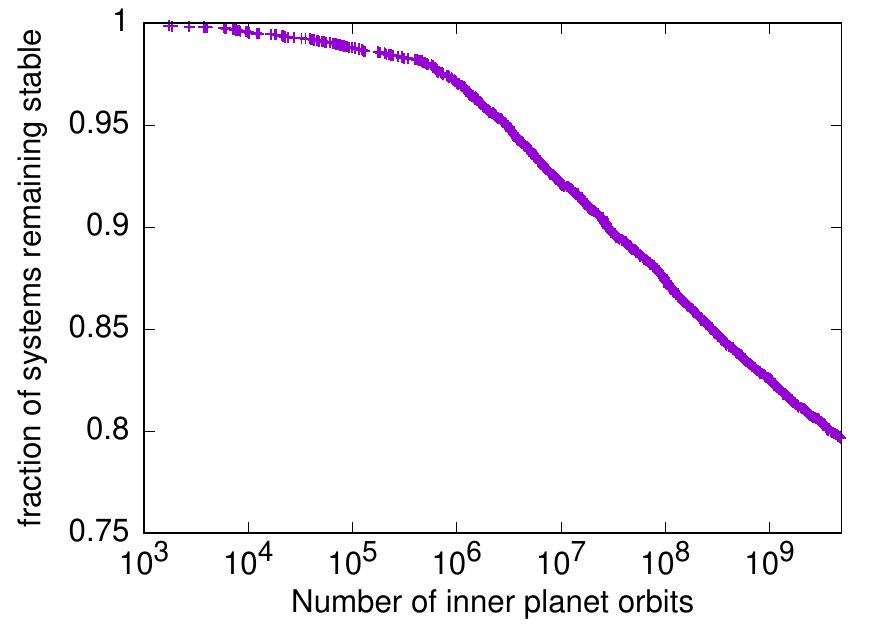}
    \caption{Fraction of simulations with no collisions as a function of the number of orbits completed by the innermost planet in the system. From $10^6-5\times10^9$ orbits, the fraction decreases steadily, indicating a flat distribution of the log of the instability timescale.
    \label{f:fraction-vs-time}}
\end{figure}
%%%%%%%%%%%%%%%%%%%%%%%%%%%%%%%%%%%%%%%%%%%

\begin{figure}
    \centering
    \includegraphics[width=3in]{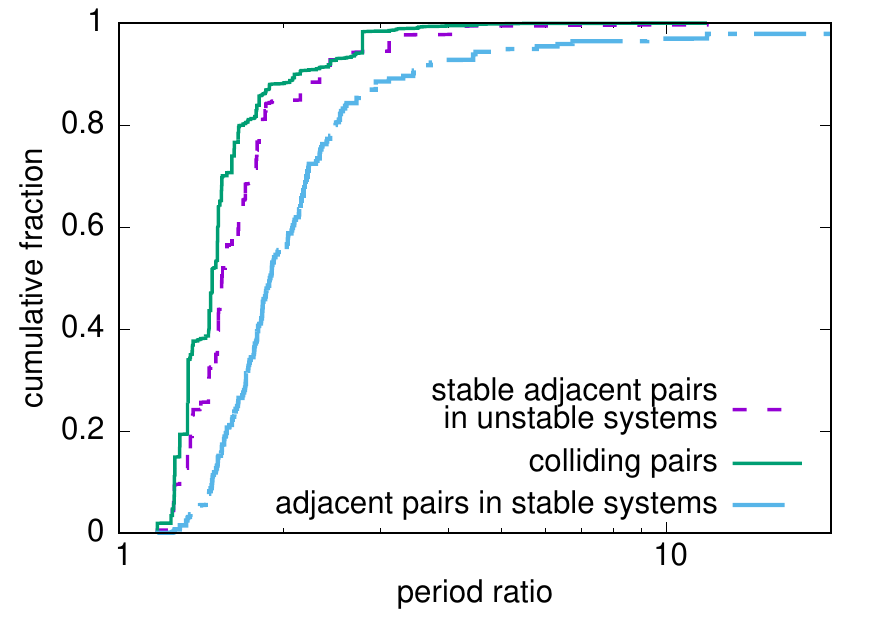}
    \includegraphics[width=3in]{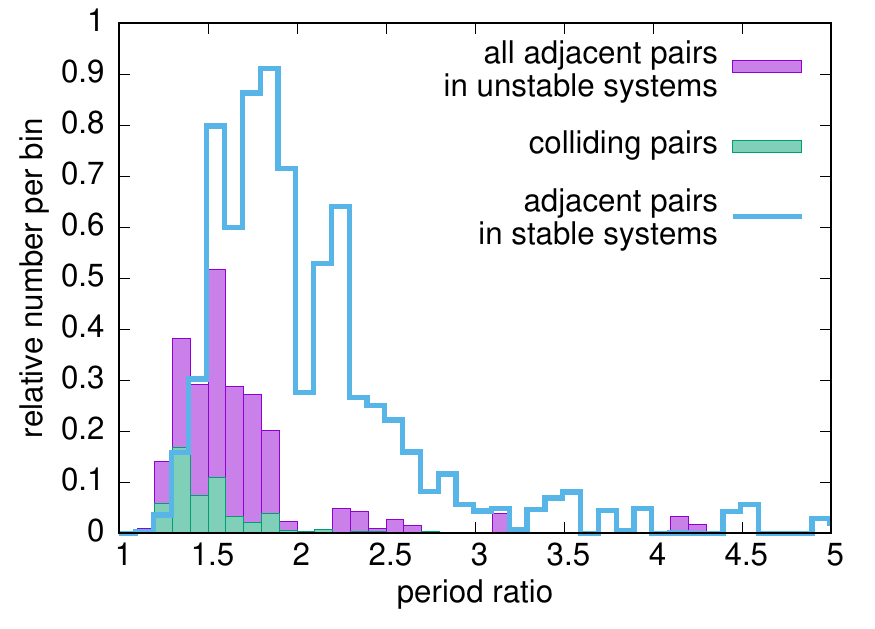}
    \caption{Period ratio distributions for adjacent planets in stable and unstable systems and for colliding planets in unstable systems. Left: Cumulative period distributions for colliding planet pairs in unstable systems (solid green line), stable adjacent planet pairs in unstable system (dashed purple line), and adjacent planet pairs in stable systems (dot-dashed blue line). Right: Histograms of all adjacent planet pairs in stable systems (open blue histogram), all adjacent planet pairs in unstable systems (filled purple histogram), and colliding planet pairs in unstable system (filled green histogram).
    \label{f:period-ratios}}
\end{figure}

Overall, $\sim20\%$ of these simulations resulted in a collision between planets in the system. 
As noted by \citet{Volk:2015}, the distribution of instability timescales is linear in log time (Figure~\ref{f:fraction-vs-time}).
We find that systems with dynamical spacings $K\lesssim12-15$ have the strongest correlation between $K$ and the probability of dynamical instability; this is consistent with the stability scaling laws discussed in Section~\ref{sec:intro}. 
However, at larger values of $K$ (where the vast majority of the observed multiplanet systems appear to lie) the relationship between stability and dynamical spacing is not clear. 
We find that, for $K\gtrsim15$, neither the fraction of surviving systems nor the instability timescale appears to be strongly correlated with dynamical spacing.
Analysis of our suite of simulations shows that the most consistent correlation between any simple dynamical parameter and the probability that the system experiences instability is with the period ratios of adjacent planet pairs.
Systems that have planet pairs with period ratios below 2 are the ones that most often display instabilities, as seen in Figure~\ref{f:period-ratios}.
In the next section, we investigate a single planetary system in detail to explore what these trends from the larger simulation set mean for the source of instabilities.

%%%%%%%%%%%%%%%%%%%%%%%%%%%%%%%%%%%%%%%%%%%
%%%%%%%%%%%%%%%%%%%%%%%%%%%%%%%%%%%%%%%%%%%

%%%%%%%%%%%%%%%%%%%%%%%%%%%%%%%%%%%%%%%%%%%
%%%%%%%%%%%%%%%%%%%%%%%%%%%%%%%%%%%%%%%%%%%
\section{Kepler-102: a case study}\label{s:casestudy}
%%%%%%%%%%%%%%%%%%%%%%%%%%%%%%%%%%%%%%%%%%%
%%%%%%%%%%%%%%%%%%%%%%%%%%%%%%%%%%%%%%%%%%%
 
 To identify the triggers of long-term instability in multi-planet systems, we investigate in detail the five-planet Kepler-102 system.
We chose Kepler-102 for three reasons: it has five confirmed planets (and thus could display rich dynamical behavior over the sampled parameters); our simulations of this system had a high incidence of instabilities; and the observed period ratios of the confirmed planets suggest mean motion resonances could be important.
Given this last point and that the probability of our simulated planetary systems experiencing an instability is not strongly correlated with dynamical spacings and only appears to be related to the period ratios of adjacent planet pairs, we first investigate whether proximity to or overlap between mean motion resonances is a viable explanation for instabilities in this system (Section~\ref{ss:MMRs}); 
however, for Kepler-102 (and later for a wider range of systems), we find that this is not a likely explanation for reasonable planet mass ranges. 
Instead, the secular structure of the simulated planetary systems appears to be much more predictive of instability. 
In Section~\ref{ss:secular-dynamics}, we describe evidence that secular chaos causes the transfer of angular momentum deficit amongst planets within the planetary system such that even initially low-eccentricity planets can be driven to orbit-crossing and collisions.

\subsection{Kepler-102 system overview}\label{ss:K102}

%%%%%%%%%%%%%%%%%%%%%%%%%%%%%%%%%%%%%%%%%%%
\begin{deluxetable*}{|c|c|c|c|c|}[hbt]
\tablecaption{Summary of the Kepler-102 system}
    \tablehead{KOI identifier & planet & orbital period (days)\tablenotemark{a} & radius ($R_{\earth}$)\tablenotemark{b}  & a (au)} 
    \startdata
    KOI-82.05 & Kepler-102 b & 5.287 & $0.50^{+0.10}_{-0.05}$ &  0.055  \\ 
    KOI-82.04 & Kepler-102 c  & 7.07 & $0.62^{+0.18}_{-0.05}$ &  0.067  \\ 
    KOI-82.02 & Kepler-102 d  & 10.31 & $1.31^{+0.06}_{-0.13}$ & 0.086 \\
    KOI-82.01 & Kepler-102 e  & 16.15 & $2.4^{+0.1}_{-0.2}$ & 0.116  \\ 
    KOI-82.03 & Kepler-102 f  & 27.45 & $0.93^{+0.11}_{-0.10}$ & 0.165  \\
    \enddata
    \tablenotetext{a}{\cite{Holczer:2016}}
    \tablenotetext{b}{\cite{Berger:2018}}
    \tablecomments{Kepler 102 system parameters retrieved from the NASA Exoplanet Archive (values taken from the indicated references). The semimajor axis values are calculated for a host star mass $M_* = 0.8 M_{\sun}$ \citep{Morton:2016}.
    \label{t:K102}}
\end{deluxetable*}
%%%%%%%%%%%%%%%%%%%%%%%%%%%%%%%%%%%%%%%%%%%

The Kepler-102 system has five confirmed planets;  three of them are of sub-Earth radius and the larger two have radii of $1.3 R_{\earth}$ and $2.4 R_{\earth}$. 
The host star's radius is $R_*=0.726^{+0.030}_{-0.028} R_{\sun}$ \citep{Berger:2018} and its mass is $M_* = 0.80\pm0.03 M_{\sun}$ \citep{Morton:2016}.
Table~\ref{t:K102} lists the observed parameters of the five planets (orbital period and radius) along with a semimajor axis value (assuming $M_*=0.8M_{\sun}$) for each planet.

In the simulations based on this system's architecture, 79 of the 100 sets of initial conditions result in a collision within $5\times10^9$ orbits; planets b and c are almost always the colliding pair despite  initially large dynamical spacings, $K\gtrsim25$. 
We identify two features of the Kepler-102 system that could be contributing to dynamical instabilities.
First is the apparent prevalence of near-commensurate orbital periods:
planets b and c are very close to a 4:3 period ratio (4:2.99), planets d and c are close to a 3:2 period ratio (3:2.06), planets b and d are close to a 2:1 period ratio (1.95:1), and planets e and f are close to a 5:3 period ratio (5:2.94).
This leads us to investigate the potential role of mean motion resonances in the dynamics of the system (Section~\ref{ss:MMRs}). 

The second notable feature of the system is that the two innermost planets are also the smallest (and therefore likely the least massive) planets in the system. 
Inward transfer of angular momentum deficit (AMD) to these small planets would produce larger increases in orbital eccentricity than for more massive planets, reminiscent of the long term instabilities that have been identified in numerical simulations of the solar system \citep{Wu:2011,Laskar:2017}.
This motivates us to investigate the role of secular chaos and AMD transfer (Section~\ref{ss:secular-dynamics}).

%%%%%%%%%%%%%%%%%%%%%%%%%%%%%%%%%%%%%%%%%%%
%%%%%%%%%%%%%%%%%%%%%%%%%%%%%%%%%%%%%%%%%%%
\subsection{Mean motion resonances in the Kepler-102 system}\label{ss:MMRs}

The proximity of the Kepler-102 planets to several first and second order resonances suggests that we investigate whether interacting mean motion resonances may contribute to dynamical chaos and instability in the system \citep[see, e.g.,][]{Mahajan:2014,Pu:2015}.
We first calculate the widths of these resonances using analytical theory and then use numerical methods to confirm the analytical estimates for the closest resonance in the system.

For nearly circular orbits, the half-width of a planet's internal first order, $(p+1):p$, MMR is given by 
\begin{equation}
\frac{\Delta P}{P} = \Big| \frac{9\alpha f_d \mu}{\sqrt{8p}} \Big|^\frac{2}{3},
\label{eq:reswidth}
\end{equation}
where $P$ is the unperturbed orbital period, $\mu$ is the mass of the planet in units of the central mass, $\alpha$ is the ratio of the exact resonant semimajor axis to the planet's semimajor axis, and the coefficient $f_d$ is a function of $p$ and $\alpha$ \citep{Malhotra:1998,Petrovich:2013}.
The coefficients $\alpha f_d$ for several internal first-order resonances are given in Table 8.5 of \cite{Murray:1999}; we list these in Table~\ref{t:reswidth} along with the expression for each resonance's half-width, $\Delta P/P$, as a function of $\mu$ given by Eq.~\ref{eq:reswidth}.

\begin{deluxetable}{|c|c|c|}
\tablecaption{Resonance half-widths for nearly circular orbits at interior first-order resonances}
    \tablehead{resonance & $\alpha f_d$ & $\Delta P/P$} 
    \startdata
    2:1 & -0.749964 & $1.786\mu^{2/3}$  \\
    3:2 & -1.54553 & $2.393\mu^{2/3}$  \\
    4:3 & -2.34472 & $2.647\mu^{2/3}$  \\
    5:4 & -3.14515 & $2.926\mu^{2/3}$  \\
    6:5 & -3.94613 & $3.159\mu^{2/3}$  \\
    \enddata
    \tablecomments{The values of $\alpha f_d$  are taken from \cite{Murray:1999}.
    \label{t:reswidth}}
\end{deluxetable}

Planets b and c are the closest to a first order MMR (the 4:3), with $(P_b - P_{4:3,c})/P_{4:3,c} = -3\times10^{-3}$.
Taking the expression for the width of planet c's internal 4:3 MMR from Table~\ref{t:reswidth}, the value of $\mu$ required for the 4:3 MMR to overlap planet b's observed orbital period is $\mu > 4\times10^{-5}$, which translates to approximately $10 M_{\earth}$. 
Given that the radii of planets b and c are each $\sim0.5 R_{\earth}$, a mass of $\sim10 M_{\earth}$ is implausibly high as it implies unphysical bulk densities of these planets,  $\rho\gtrsim400$~g~cm$^{-3}$~\citep{Fortney:2007}. We conclude that for realistic bulk densities, planets b and c are of sufficiently low mass that the 4:3 resonance cannot be the direct source of dynamical instability.

Given that planet b's observed orbital period is in-between planet c's internal 4:3 MMR and planet d's internal 2:1 MMR, we can also  estimate the planet masses that would be required for resonance overlap to significantly affect planet b's orbital evolution.
The fractional separation of the locations of these two MMRs on either side of planet b is
$(P_{4:3,c} - P_{2:1,d})/{P_b} = 0.0279$.
Dynamical chaos and instability due to MMR overlap would occur if the sum of the half-widths of the two neighboring MMRs exceeded their separation. 
For the resonances above, this requires $\mu > 5\times10^{-4}$, i.e., planet masses larger than $13 M_{\earth}$.
Planet d is larger than planets b and c, but at $~\sim1.3 R_{\earth}$, the large mass required for resonance overlap is still not plausible because it would imply an unphysical bulk density, $\rho\gtrsim30$~g/cc.

For moderately eccentric orbits, we can estimate the widths of these MMRs using the analytical expressions based on the pendulum model for MMRs 
\citep[][see Appendix~\ref{a:res}]{Murray:1999}.
If we again consider the proximate 4:3 MMR between planets b and c, we can calculate the eccentricity at which planet c's interior 4:3 resonance width will encompass planet b's observed orbital period. 
If we assume the mass of planet c is $0.35M_{\earth}$ (which represents a high-end mass limit requiring a density 1.5 times larger than the Earth's), planet b would need an eccentricity of 0.25 for the resonance width to reach its observed orbital period; this is slightly larger than the eccentricity at which planet b would cross planet c's semimajor axis. 
Given that the {\it Kepler} multiplanet systems have been shown to typically have much smaller eccentricities than this, it is again clear that planets b and c are not actually in resonance.

\begin{figure}[hb]
    \centering
    \includegraphics[width=3.5in]{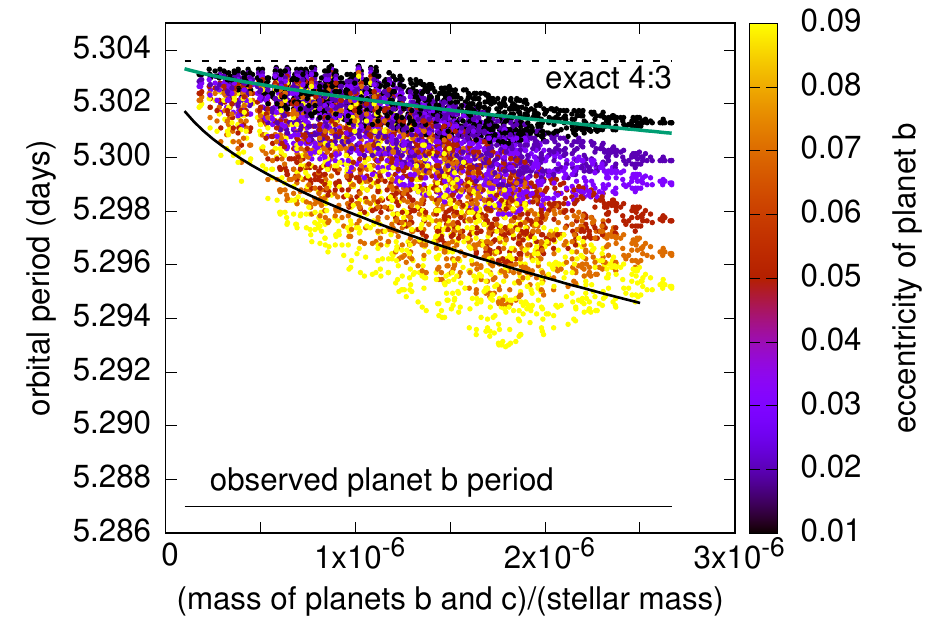}
    \caption{
    The maximum half-width of orbital period variations due to the 4:3 resonance between Kepler-102 planets b and c as a function of the combined masses of planets b and c (in units of the host star's mass) at various eccentricities. The results from numerical simulations are shown with the colored points (see color bar for eccentricity values).  The analytical estimate of the resonance widths at  zero eccentricity (equation~\ref{eq:reswidth}) and at higher eccentricity ($e=0.09$, see Appendix~\ref{a:res}) are shown with the green and grey curves, respectively; these agree well with the numerical simulation results. We see that, for moderate eccentricities ($e<0.1$) and a plausible range of planet masses, planet b's orbital period is well outside the 4:3 resonance with planet c.
    \label{f:4-3-resonance}
    }
\end{figure}

These analytical estimates of resonance width consider only the mass of each planet in isolation. To take account of both planets' masses, an empirical rule-of-thumb is to interpret $\mu$ in Table~\ref{t:reswidth} as the sum of the planet masses in units of the stellar mass (see, e.g., \citealt{Deck:2013}, who showed that for two massive planets, resonance widths are much more sensitive to the sum of their masses than to their mass ratio). The resulting resonance width estimates as a function of $\mu$ are shown in Figure~\ref{f:4-3-resonance} for the case of nearly circular orbits as well as for orbital eccentricity 0.09.
We confirm these analytical estimates with numerical techniques to compute the widths of resonances for plausible planet masses (densities). Following \cite{Wang:2017}, we do this by constructing surfaces of section for the 4:3 resonance between planets b and c (in the co-planar case) as follows. 
First, planets b and c are randomly assigned masses from the \cite{Wolfgang:2016} mass-radius relationship (we ignore the other planets for this calculation). 
Planet c is given a circular orbit, and its initial position relative to the star defines the x-axis. 
Then, for each pair of masses for planets b and c, a sequence of separate simulations are performed. 
Planet b is sequentially assigned an initial eccentricity from the list  $e=[0.01,0.02,0.03,0.05,0.07,0.09]$. 
For each eccentricity value, planet b's perihelion is chosen to be at an angle $\phi$ away from the location of planet c (i.e., away from the x-axis); the angles $\phi =0-90^\circ$ are explored via separate simulations in $3^\circ$ increments (this range in initial $\phi$ ensures the full resonant phase space is simulated). 
For each $e$ and $\phi$ combination, planet b's semimajor axis is initialized at the exact 4:3 resonant value (assuming a stellar mass of $M_* = 0.8 M_{\sun}$), and 2000 orbits of planet b are simulated (starting planet b at perihelion). 
Every time planet b passes through perihelion, its semimajor axis and angular separation from planet c are calculated and recorded. 
These surfaces of section for the exact resonant orbit are used to determine the largest width in semimajor axis achieved by libration within the 4:3 resonance with planet c. These numerically computed estimates of resonance width are shown in Figure~\ref{f:4-3-resonance} in terms of the maximum variations in planet b's orbital period as a function of the combined total mass of planets b and c.
It is clear that planet b's observed orbital period does not lie within the resonance for reasonable planet mass and eccentricity combinations.

\begin{figure*}
\begin{minipage}{0.5\linewidth}
\begin{minipage}{0.9\linewidth} 
\caption{An overview of the locations and widths of mean motion resonances in the Kepler-102 planetary system. 
Orbital periods are on the y-axis (note the y-axis is discontinuous and the scaling is not the same in each region) and eccentricity is on the x-axis. 
The location of each transiting planet in orbital period is indicated by a circle (with relative sizes reflecting planetary radii) and a solid horizontal line that extends to the eccentricity at which that planet would cross another planet's orbit (period uncertainties are smaller than the line width). 
The locations of a subset of each planet's interior and exterior resonances (up to 4th order) are indicated by horizontal dashed lines with widths shaded (the color of each resonance matches its planet). 
The mean motion resonances are labeled such that ``4-b:3'' means a particle at that location would complete 3 orbits in the same amount of time that Kepler-102 b takes to complete 4 orbits.
The darker shaded region is the resonance width (in the test-particle limit; Appendix~\ref{a:res}) for a low-end planet mass estimate, and the lighter shaded region is for a high-end mass estimate. 
For Kepler-102 b, c, d, and f, the low-end mass limit represents a planet with a density equal to $0.5\rho_{\earth}$, and the high-end limit is a density of $1.5\rho_{\earth}$. 
For Kepler-102 e, the 1--$\sigma$ limits on the mass from RV measurements \citep{Marcy:2014} were taken.
We assumed a stellar mass of $0.8M_{\sun}$ when calculating the resonance widths.
\label{fig:K102-res}}
\end{minipage}
\end{minipage}
\begin{minipage}{0.5\linewidth}
    \includegraphics[width=0.8\linewidth]{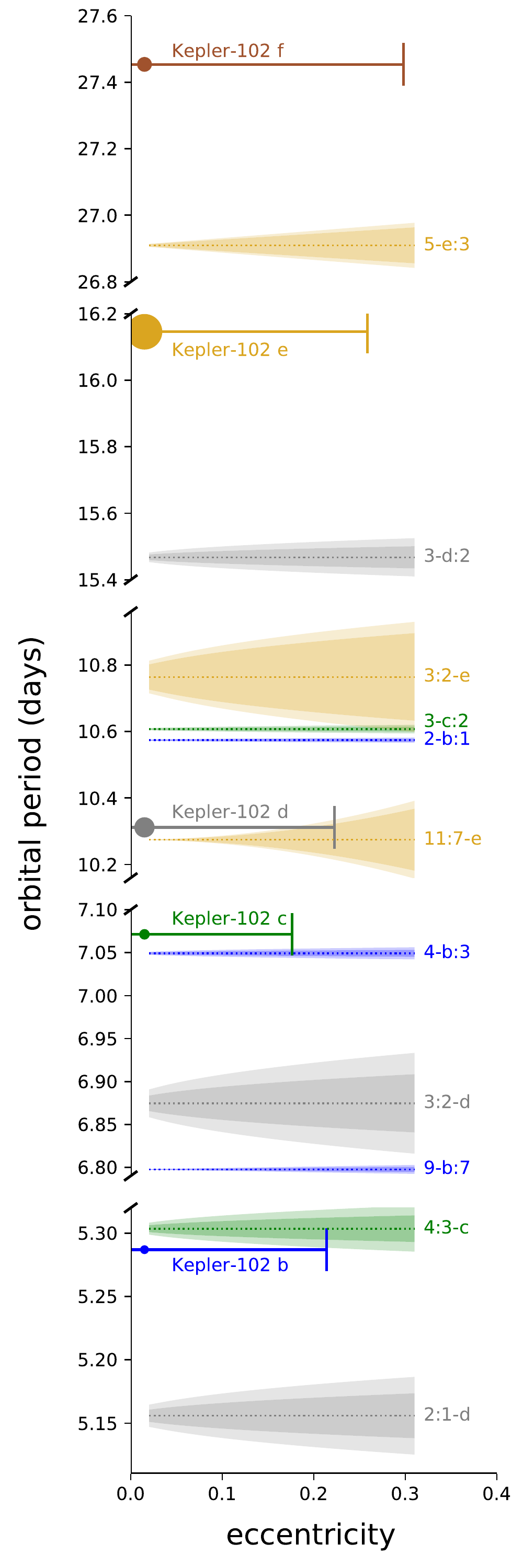}
\end{minipage}
\end{figure*}

The analytical estimates of resonance widths for near-circular orbits (Eq~\ref{eq:reswidth}) and for higher-eccentricity orbits (Appendix~\ref{a:res}) agree quite well with the results of numerical simulations in Figure~\ref{f:4-3-resonance} at low and high eccentricity, respectively.
This confirms that the analytical expressions are sufficient to check for the overlap of MMRs in the typically low-eccentricity {\it Kepler} systems as well as for checking which individual MMRs are close enough to observed planet periods to plausibly be dynamically relevant.
We use the expressions in Appendix~\ref{a:res} to calculate the widths of the first, second, third, and fourth order resonances for each confirmed planet in the Kepler-102 system.  
For planets b, c, d, and f, we take a range of planet masses corresponding to a range planet densities 50\%--150\%  Earth's density; for the largest planet, planet e, we take the 1--$\sigma$ limits on its mass ($6-9 M_{\earth}$) from RV measurements \citep{Marcy:2014}.  
With these planet mass ranges, we calculated the resonance widths for eccentricities in the range 0.02--0.3. Figure~\ref{fig:K102-res} shows the results for a subset of resonances in the vicinity of each observed planet in the Kepler-102 system. 
It is clear from these calculations that at the low eccentricities typical of {\it Kepler} systems, all of the Kepler-102 planets are reasonably well-separated from their mutual MMRs.
Even at eccentricities as large as $e\sim0.3$, there is no overlap of neighboring mean motion resonances to directly cause instabilities. 
Furthermore, we can rule out the possibility that planets b and c could be in stable libration within the 4:3 resonance: planet b's observed orbital period is too far from the 4:3 MMR for this to be possible. 
However, it is also evident that planet b is the most vulnerable to mean motion resonant perturbations, needing only a moderate eccentricity excitation, $e\gtrsim 0.1$, to reach the boundary of planet c's interior 4:3 resonance.
In the next section, we explore how such eccentricity excitation might occur.

Before moving on to the secular dynamics of the Kepler-102 system, we note that the relative proximity of MMRs to planets in the Kepler-102 system raises the possibility of using transit timing variations (TTVs) to constrain the masses of some of the system's planets.
TTV measurements have been reported for the four outer planets (c, d, e, and f) in the Kepler-102 system based on the long-cadence {\it Kepler} data. 
\cite{Hadden:2014} used TTVs to estimate the mass of Kepler-102 d based on the assumption that the observed TTVs in the system for planets c, d, and e were the result of the near 3:2 MMRs between planets c and d and planets d and e.
However this mass estimate is unfortunately likely erroneous in light of subsequent data; a more recent measurement and analysis of TTVs from all available long-cadence {\it Kepler} data \citep{Holczer:2016} does not indicate statistically significant periodicity in the measured TTVs for Kepler-102.
Additionally, transits of Kepler-102 b were not part of the dataset analyzed by \cite{Hadden:2014}, so its near resonance with planet c was not considered as a possible source of TTVs.
In Appendix~\ref{a:ttvs}, we present a brief analysis of how the near-resonances in the Kepler-102 system could induce TTVs, though the \cite{Holczer:2016} transit times do not show observational evidence of them.

\subsection{Secular dynamics in the Kepler-102 system}\label{ss:secular-dynamics}

We have shown that, at low to moderate eccentricities, MMRs are not a likely direct source of instabilities in our long-term simulations based on the Kepler-102 system. 
Here we turn our attention to the secular structure and evolution of the system.
For each set of assumed Kepler-102 planet masses, we use the Laplace-Lagrange linear secular theory \citep{Murray:1999} to calculate the basic secular architecture of the system.
Briefly, this theory assumes that the planets in the system can be modeled as rings, with the mass of each planet spread out along its orbit. The shapes and orientations of the rings change slowly with time under the mutual gravitational perturbations of the planets. In the linearized secular approximation, the time variation of the planets' eccentricities is decoupled from that of their mutual inclinations.  The time evolution of the eccentricity vector of each planet is expressed as a superposition of linear modes (``secular modes'') whose frequencies depend only on the masses and orbital periods of the planets and on the mass of the host star. 

Because the secular frequencies depend on planet masses, and the planet masses are not known, there is a wide variety of possible secular architectures for the Kepler-102 system. 
We calculated the linear secular solution for this system by randomly sampling the full range of possible planet masses and initial conditions. Our calculations find a few general properties of note.
One is that it is not uncommon for two of the  eccentricity mode frequencies to be of similar magnitude and for these two modes to have roughly equal power in the secular solution for the inner planets' eccentricity vectors. This near-degeneracy of a pair of secular modes means that the phenomenon of mode beating can occur and can lead to large eccentricities for some planets on secular timescales.
This phenomenon explains the very shortest instability timescales found in a few of our simulations, representing initial conditions that simply lead to constructive combinations of secular mode amplitudes that causes planet-crossing and destabilizing close encounters. In the majority of our simulations, however, it is the slower chaotic transfer of AMD amongst the planets which eventually builds up the eccentricity of the inner, low mass planet. 
We note that the orbit of planet b is close to the location where a secular resonance would occur in the test-particle approximation (where planet b's mass is zero); in these cases, the precession of a test particle's orbit at the location of planet b would nearly match one of the four eccentricity frequencies induced by planets c-f, which would result in a large forced eccentricity for the test particle's orbit. 
This is interesting because planet b is the smallest planet in the system; if planet b is a particularly low-density planet with a mass much smaller than the other planets in the system, then it could be subject to more significant secular eccentricity variations.
Thus it is plausible that the instabilities in the simulated Kepler-102 systems are driven by secular interactions.

 In the secular (orbit-averaged) approximation, the semi-major axes of the planets remain constant over time, and the total angular momentum deficit (AMD) of a planetary system is conserved~\citep[see, e.g.,][]{Laskar:2017}; the total AMD of an N-planet system is given by the sum of the AMDs of each planet:
\begin{equation} \label{eq:amd}
\begin{split}
& AMD_{total} = \sum_{j=1}^{N} AMD_j \\
& AMD_j =  \\
& \frac{m_j M_*}{m_j + M_*} \sqrt{G(m_j + M_*) a_j} \left (1 - \sqrt{1-e_j^2} \cos i_j \right), \\
\end{split}
\end{equation}
where $G$ is the universal constant of gravitation, $M_* $ is the mass of the star, and $m_j$, $e_j$, and $i_j$ are the mass, eccentricity, and inclination of the $j$th planet.
In the linear approximation, the eccentricities and mutual inclinations have quasi-periodic time variation, with maxima given by the constructive interference of all the linear secular modes~\citep{Murray:1999}. 
Going beyond linear secular theory, \cite{Lithwick:2011} have suggested that when there are near-commensurate secular frequencies in a planetary system, higher order secular perturbations can cause `secular chaos', leading to changes in the secular modes and net transfer of AMD between planets in the system that is distinct from the regular secular oscillations. 
Through secular chaos, it is possible over long timescales for the total AMD of a planetary system to be transferred almost fully to the shortest period planet in the system \citep{Wu:2011,Petrovich:2019}.
Given our findings above from a simple analysis of the range of secular structures for Kepler-102 (the tendency to have nearly commensurate mode frequencies and the tendency for planet b's orbit to lie near where a massless particle's free precession would match one of these modes), secular chaos seems likely to occur in this system.
Should such AMD transfer be realized, the innermost planet's eccentricity could be excited to high values, potentially leading to close encounters with the next neighboring planet and thereby triggering a strong dynamical instability; alternatively, eccentricity excitation could increase perturbations from near-resonances and trigger further eccentricity excitation and planet-planet encounters.
An eccentricity $e\gtrsim0.2$ for planet b places it on a planet-crossing orbit with its neighbor, planet c. In reality, planet b is vulnerable to instability at even lower eccentricity, $e\sim0.1$ because such an eccentricity places it close to the boundary of the 4:3 resonance with planet c (see Section~\ref{ss:MMRs} and Figure~\ref{fig:K102-res}).

This motivates us to examine how the masses and eccentricities in the Kepler-102 system affect the maximum possible eccentricity excitation for planet b in the limit where the total initial AMD of the system is transferred inward to planet b.
To test and to illustrate this conjecture of `AMD transfer by secular chaos', we carry out a simple numerical experiment.
We assume that the Kepler-102 system is exactly co-planar. 
We assign masses for planets b, c, d, and f from the mass-radius statistical relationship of \cite{Wolfgang:2016} and we assign a mass for planet e from the RV mass limits \citep{Marcy:2014}.
We assume that planet b starts on a circular orbit, and that the other 4 planets all start with identical eccentricities. 
We assign common initial eccentricities for planets c-f in the range $0.01-0.09$, calculate the total AMD of the system, and then calculate the maximum possible eccentricity of planet b for that total AMD.

\begin{figure}[htb]
        \includegraphics[width=3.25in]{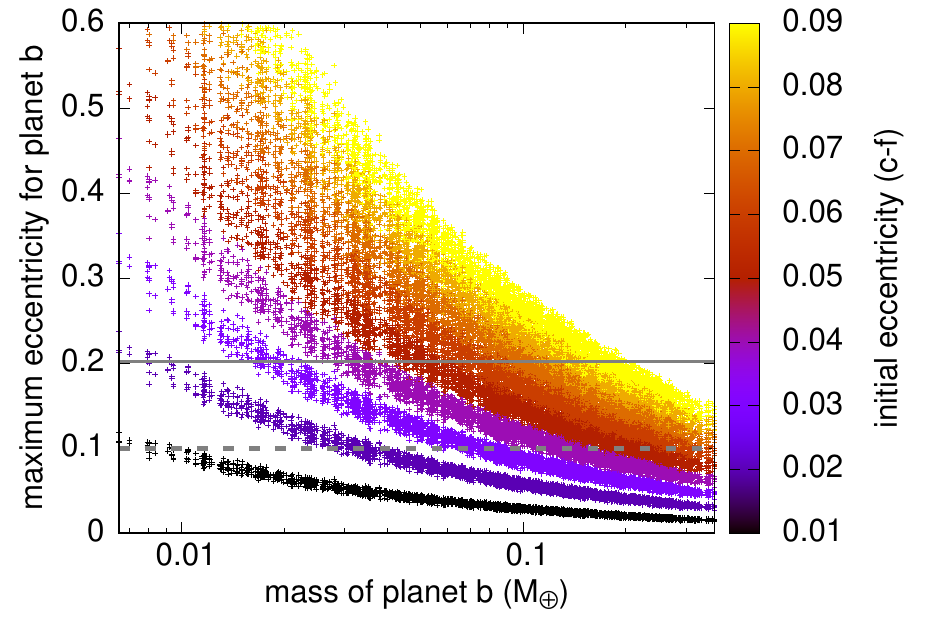} 
    \caption{The maximum eccentricity Kepler-102 b can achieve assuming it receives all the AMD of the other 4 planets' orbits as a function of planet b's mass. The colors indicate the initial eccentricity given to planets c-f (each planet has the same initial eccentricity). The solid gray horizontal line is the eccentricity at which planet b crosses planet c's semimajor axis, a strict limit for stability of the system; the dashed gray horizontal line indicates the eccentricity at which the near mutual 4:3 MMR between planets b and c becomes a likely source of instability.
    \label{f:amd-example}}
\end{figure}

Fig~\ref{f:amd-example} shows the results of this numerical experiment. It is evident that planet b's eccentricity can be excited to large values, $e\gtrsim0.2$, even in cases where the initial eccentricities of the other planets are very small, $e\sim0.02-0.03$.  
We also observe from Fig.~\ref{f:amd-example} that, unsurprisingly, the lower the mass of planet b the more vulnerable it is to instability by AMD transfer.
Considering that the eccentricities of planets in multi-planet systems detected by Kepler are typically $0.02-0.04$ \citep[see, e.g.,][]{Hadden:2014,Xie:2016}, the results of our numerical experiment show that in order for planet b's eccentricity to not be able to exceed $\sim0.1$ via inward AMD transfer,
planet b's mass must exceed $\sim0.1~M_{\earth}$.
Considering the uncertainty in the observed radius of planet b, this lower limit for planet b's mass corresponds to a bulk density in the range $(0.46-1.1)$ times Earth's bulk density. For comparison, we note that the mass-radius statistical relationship of \cite{Wolfgang:2016} gives planet b's likely mass range to be $0.007-0.35~M_{\earth}$. 
It is possible that there are additional as-yet undetected planets at larger orbital periods in the Kepler-102 system not included in our analysis. However, any additional planets would only strengthen the mass limit on Kepler-102 b as they would provide additional AMD to the system that could be subject to inward transfer.

Examining the results of our long term simulations, we find that simulations in which planet b exceeds the mass threshold of $0.1~M_{\earth}$ are almost twice as likely to remain stable for $5\times10^9$ orbits of planet b; 16\% of the simulations below this threshold were stable compared to 28\% of the simulations above this threshold. 
Unsurprisingly, many of the unstable simulations also exceed this mass threshold.
Our simple numerical experiment does not account for a distribution of initial eccentricities for the other planets in the system and does not consider the possibility that the other small inner planet, planet c, is also likely subject to some eccentricity excitation by inward transfer of AMD.
We also note that while some of our simulations with planet b's mass below $0.1~M_{\earth}$ remained stable for $5\times10^9$ orbits, this timespan is only a small fraction of the real observed Kepler-102 system's likely gigayear age; it is likely that  these system architectures would become unstable due to AMD transfer if allowed to evolve an order of magnitude longer in time.
We conclude that this approximate lower limit for planet b's mass is probably necessary but not sufficient for long term stability.

\begin{figure}
    \centering
    \includegraphics[width=3in]{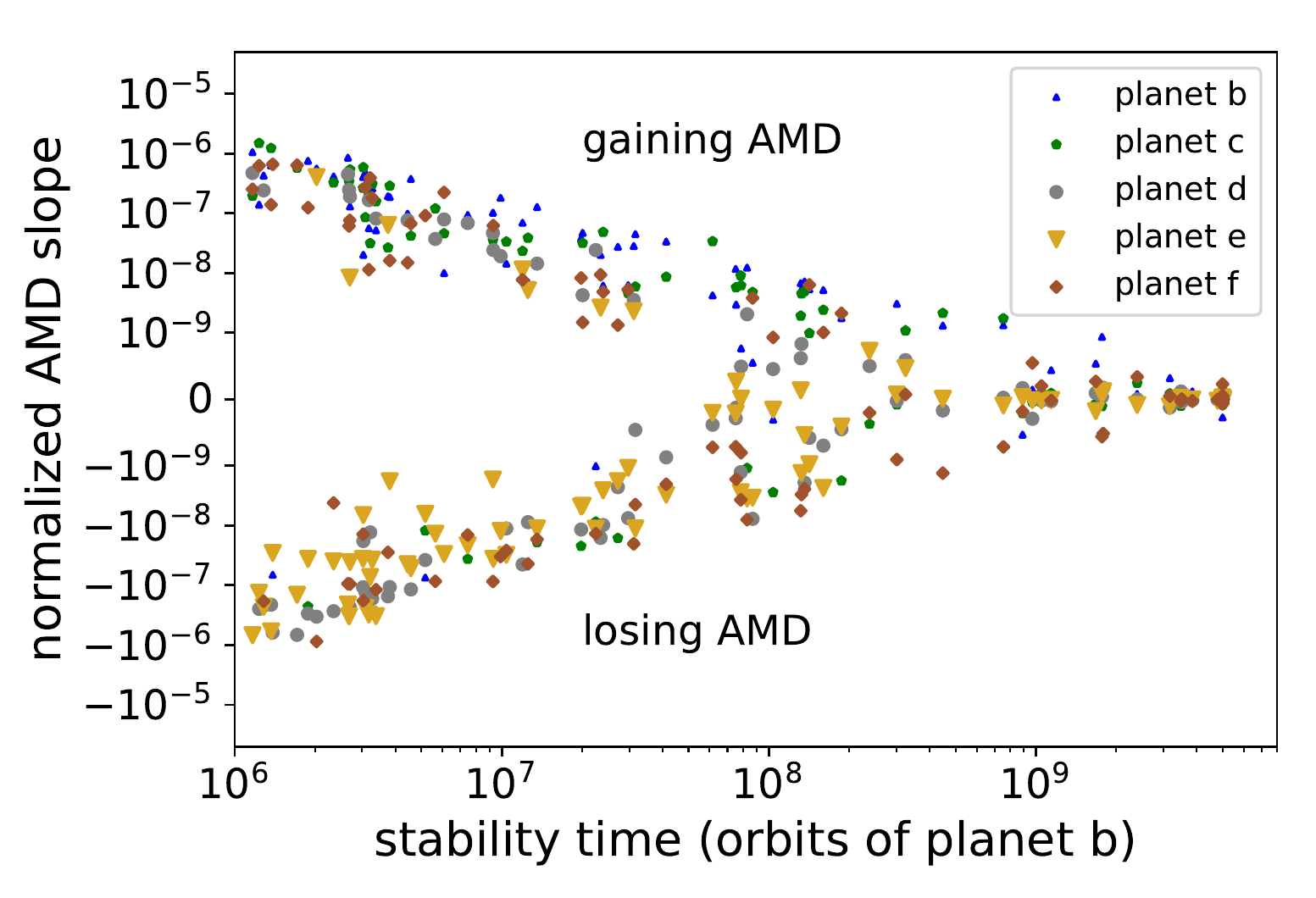}
    \caption{Slopes from a linear fit to each simulated planet's normalized AMD vs stability timescale for that simulation. The slope is in units of fractional AMD lost/gained per orbit of planet b. The innermost planets (b and c) most often have positive AMD slopes, indicating net gains in AMD. The most massive planet (e) most often has a negative AMD slope, indicating a net loss of AMD. }
    \label{f:amd-slope}
\end{figure}

The above analysis has shown that the inward transfer of even modest amounts of AMD has the potential to destabilize the Kepler-102 system.
To see if this does, in fact, occur in the N-body numerical simulations, we analyzed the time-evolution of each planet's AMD in each simulation. 
To quantify the loss or gain in each planet's AMD we calculated a best-fit linear slope to its normalized AMD over the simulation time (the units of this slope are fractional change in AMD per orbit of planet b). 
For the unstable simulations, we excluded the end portion of the simulation during which planets evolved into crossing orbits and were thus not dominated by secular interactions.
Figure~\ref{f:amd-slope} shows this normalized AMD slope for each planet as a function of how long that system remained stable.
The systems that went unstable earliest have the largest AMD slopes, while all the stable systems (at $5\times10^9$ orbits of planet b) have very small slopes. 
Visual inspection of the AMD evolution of each planet in a subset of the moderately long-lived unstable simulations confirms that these simple linear slope fits are indeed reflecting long-term trends in the AMD evolution.
For the unstable systems, the innermost, smaller planets (planet b and c) tend to have positive slopes, while the most massive outer planet (planet e) tends to have a negative slope.
This means that most of the unstable systems experienced a net inward transfer of AMD, consistent with the idea that secular chaos contributes to instabilities in our simulations of the Kepler-102 system.

The results shown in Figure~\ref{f:amd-slope} are based on the cpu-intensive, long-term simulations of the Kepler-102 system. 
In the next section, we discuss how short simulations of planetary systems can reveal which systems are most prone to destabilizing transfers of AMD.

%%%%%%%%%%%%%%%%%%%%%%%%%%%%%%%%%%%%%%%%%%%
%%%%%%%%%%%%%%%%%%%%%%%%%%%%%%%%%%%%%%%%%%%
\section{Predicting long term (in)stability}\label{s:spectra}
%%%%%%%%%%%%%%%%%%%%%%%%%%%%%%%%%%%%%%%%%%%
%%%%%%%%%%%%%%%%%%%%%%%%%%%%%%%%%%%%%%%%%%%

The analysis of the Kepler-102 system in the previous section has provided useful insights into possible drivers of dynamical instability, but does not provide a way to predict long term stability. 
Here we use the results of our large suite of long integrations supplemented with a suite of short integrations of the same systems, to identify a tool to predict stability/instability from the short simulations. 
This tool is based on the concept of spectral entropy of conservative dynamical systems introduced in \cite{Noid:1977} and \cite{Powell:1979}. Briefly, the fast Fourier transform of the time series from a numerical integration of a dynamical system is dominated by a small number of frequency components if the system is regular (stable), but it has many weaker frequency components if the system is chaotic (unstable); thus, the number of significant frequencies (``spectral number") in the power spectrum is diagnostic of (in)stability.
This method has been recently used to investigate orbital stability and chaos of stellar orbits in the Galaxy~\citep[][and references therein]{Lepine:2017} and to investigate regions of regular and chaotic orbital motion in a few individual exoplanet systems \citep{TadeudosSantos:2012,Alves:2016}. 
In the context of the solar system, \cite{Michtchenko:2002} employed this method to diagnose chaotic diffusion of the asteroid 1459 Magnya and its collisional family, and \cite{Kotoulas:2004} used this method to map chaotic regions in the vicinity of Neptune's mean motion resonances in the Kuiper belt.
Below, we employ this method to demonstrate that the power spectrum of the time series of the planets' AMD, eccentricity, and inclination from short integrations can reveal whether a system is long term stable.

For each of the multi-planet {\it Kepler} and {\it K2}  systems that we simulated in our long-term integration set (Section~\ref{s:keplermultis}), we re-integrated each set of initial conditions over $5\times10^6$ initial orbital periods of the innermost planet, recording the orbital elements and AMD of each planet at 3000 evenly spaced points. 
This time span is sufficiently long to capture multiple cycles of the longest-period linear secular modes in the vast majority of these systems.
We then performed a fast Fourier transform, FFT (using the numpy fft package in python), of the time series of the semimajor axis, inclination, eccentricity, and AMD of each planet in each simulation. 
For an input time series with 3000 points evenly spaced in time, this yields an estimate of the power associated with 1500 evenly linearly spaced frequencies where the highest frequency is twice the sampling frequency and the total length of the time series determines the lowest frequency.
 Following \citet{Michtchenko:2002}, we characterize the results of the FFT by considering how many frequencies in the FFT have  power $\geq5\%$ of the peak frequency in the power spectrum.
While \citet{Michtchenko:2002} considered the absolute number of frequencies above this threshold (spectral number) as their parameter of interest, here we make a slight modification to instead define a ``spectral fraction", i.e., the fraction of the frequencies in the FFT that exceed the $5\%$ threshold; this ensures our parameterization does not depend on the length or cadence of the input time series (because the number of entries determines the number of frequencies in the FFT).

\begin{figure*}
    \centering
    \begin{tabular}{cc}
    \includegraphics[width=3.3in]{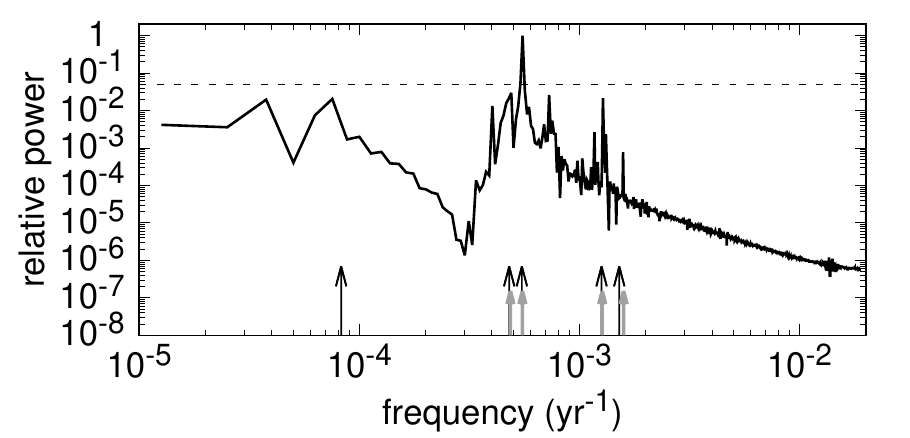}
    \includegraphics[width=3.3in]{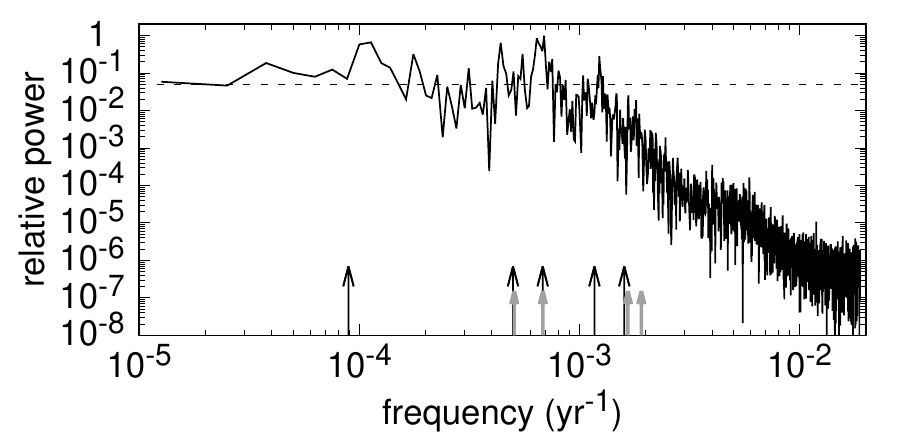}
    \end{tabular}
    \caption{Power spectra for the AMD of the innermost planet in two different versions of the Kepler-102 system; these spectra are obtained from short integrations, just $5\times10^6$ orbits. From long integrations ($5\times10^9$ orbits) of the same systems, we know that the case shown in the left panel is long term stable whereas the case in the right panel is long term unstable. 
    In each case the five eccentricity secular frequencies predicted by linear secular theory are indicated for reference as the tall black arrows and the four inclination secular frequencies are shorter gray arrows. Each power spectrum has been normalized to the power in the most dominant frequency, and the dashed horizontal line indicates 5\% of this power.
    \label{f:spectra}}
\end{figure*}

Figure~\ref{f:spectra} shows the FFT of the AMD of planet b in two versions of the Kepler-102 system. 
In the longer integrations, the case shown in the left panel survived the full $5\times10^9$ orbits, while the case on the right did not.
In the short integration, the FFT of the stable case has only a few frequencies that rise above the 5\% of the peak power threshold and its spectral fraction is $2\times10^{-3}$; these few frequencies occur in a well defined single peak in the FFT close to one of the linear secular modes of this system. However, in the short integration of the unstable case there are a significantly larger number of spectral frequencies above the 5\% threshold, and its spectral fraction is $4.5\times10^{-2}$. We see that these two cases have qualitatively different power spectra in the short integrations, and the spectral fraction is small for the stable case and larger for the unstable case.

We note that small spectral fractions for the AMD evolution of a planet is what we would expect in cases where linear secular theory accurately describes the evolution of a planet's orbit; i.e., when the eccentricity and inclination evolution are decoupled and can be described as a sum of a limited number of secular frequencies. 
When this is the case, the AMD evolution (which reflects a combination of the eccentricity and inclination evolution, see equation~\ref{eq:amd}) will have approximately the same number of dominant frequencies that are similar in magnitude to the secular frequencies.
For reference, we show the predicted linear secular frequencies for our simulated Kepler-102 systems in Figure~\ref{f:spectra}.
In the stable system, we can see that there are relatively few local peaks in planet b's AMD power spectrum and that they occur at frequencies similar to the range of the secular frequencies; the same cannot be said of the unstable system's AMD power spectrum for planet b.

\begin{figure*}
    \centering
    \includegraphics[width=6in]{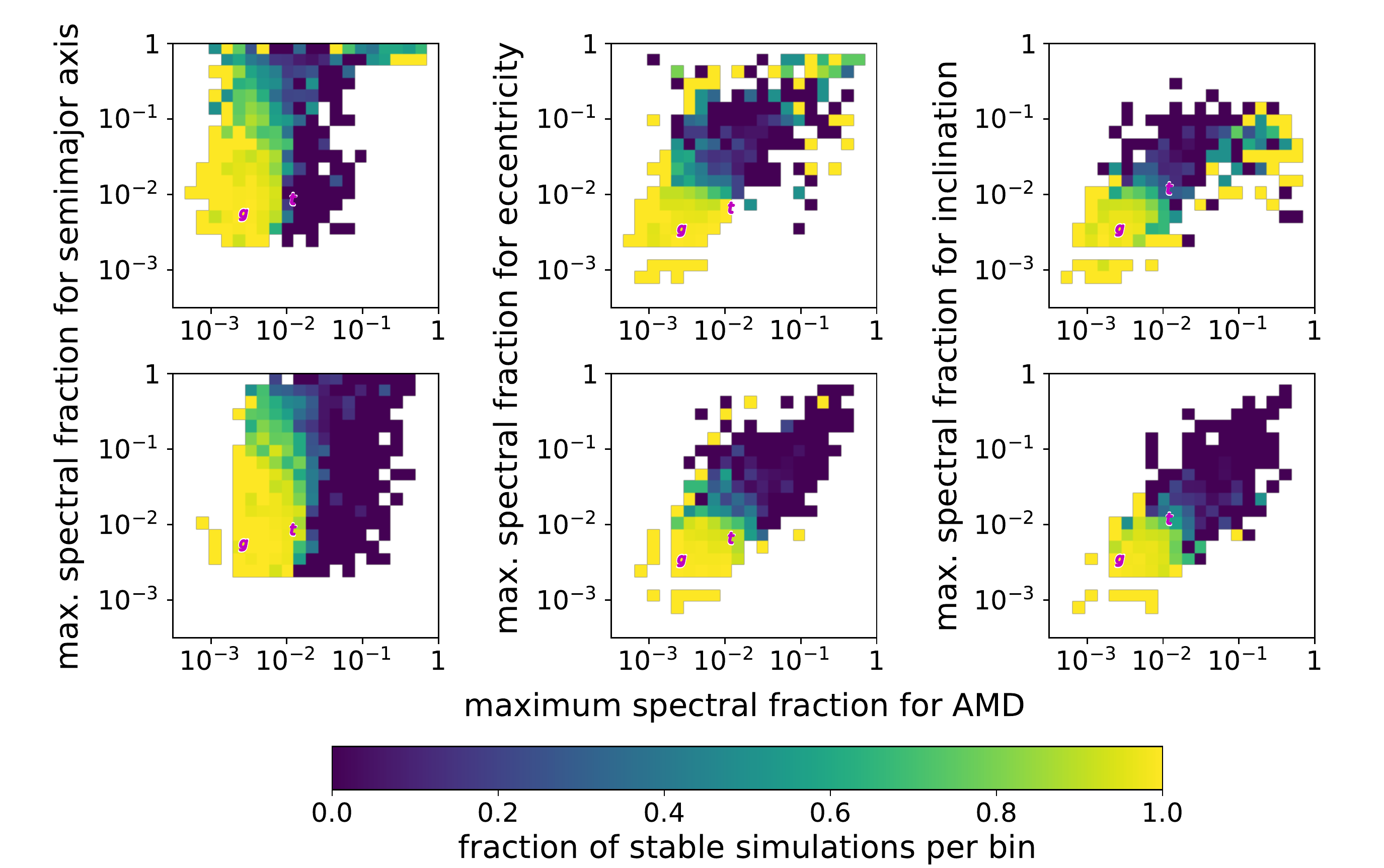}
    \caption{
    The fraction of systems that remain stable at $5\times10^9$ orbits of the innermost planet (binned color map) as a function of each system's maximum spectral fractions computed from short integrations.
    The x-axes show the maximum spectral fraction calculated from the AMD time series of each system's planets. The y-axes display the maximum spectral fraction from the planets' semimajor axis time series (left panels), eccentricity time series (middle panels), and inclination time series (right panels).
    The top row of panels is for all of our simulated cases of {\it Kepler} and {\it K2} multi-planet systems with four or more planets; the bottom row of panels excludes the subset of systems that have planets near/in first order mean motion resonances.
   \label{fig:spectral-fractions}}
\end{figure*}

We calculated the spectral fractions based on short integrations for all of the planets in all of the simulated {\it Kepler} and {\it K2} systems to assess how well they predict stability or instability in the longer simulations. 
We exclude from our analysis cases that went unstable on timescales shorter than $3\times10^6$ orbits as the power spectra from such short simulations will not include some of the longer period secular modes (such short-lived systems also reveal themselves with very little cpu time and thus do not need predictors of instability).
How system stability in our long simulations depends on the spectral fractions from the short simulations is summarized in Figure~\ref{fig:spectral-fractions}.
To construct each panel of Figure~\ref{fig:spectral-fractions}, we binned our simulations according to the largest spectral fraction from the AMD time series of any individual planet in the system (x axes) and the largest spectral fraction for the semimajor axis time series (y-axes; left panels),  the eccentricity time series (y-axes; middle panels), and the inclination time series (y-axes; right panels). 
The color assigned to each spectral fraction bin indicates the fraction of systems in the bin that survived for $5\times10^9$ orbits of the innermost planet in our long simulations. 
(Using the average instead of the largest spectral fraction yields very similar plots.)

For reference, we also ran a numerical integration of the solar system's eight major planets and computed the spectral fractions for the giant planets (calculated over a timespan equal to $5\times10^6$ Jupiter orbits) and terrestrial planets (calculated over a timespan equal to $5\times10^6$ Mercury orbits); the maximum spectral fractions for the giant planets and for the terrestrial planets are indicated by `g' and `t', respectively, in Figure~\ref{fig:spectral-fractions}.
This is a useful comparison because long-term simulations of the giant planets have shown their orbits to be stable (even in cases where their maximum Lyapunov exponents indicate chaotic behavior, see \citealt{Hayes:2008}), whereas the current orbits of the terrestrial planets do allow for potential future collisional trajectories \citep{Laskar:2009} that might be attributable to secular chaos \citep{Lithwick:2014}.

The top panels of Figure~\ref{fig:spectral-fractions} display the results for the entire set of planetary systems described in Section~\ref{s:keplermultis}, and it is immediately apparent that the stable planetary systems are not randomly distributed in spectral fraction. 
The stable systems are concentrated at small AMD spectral fractions; they also tend toward low spectral fractions in eccentricity and inclination  (which are both also represented in the AMD analysis).
There is not a strong trend with the spectral fraction of the semimajor axis; this is unsurprising as non-resonant planets should not have strong features in their semimajor axis power spectra.

Nevertheless, there are a few stable systems in the top panels of Figure~\ref{fig:spectral-fractions} that do not have small spectral fractions in AMD, $e$, and/or $i$. 
Conjecturing that these outliers are systems that are  affected by mean motion resonances,
we took our set of simulation initial conditions and determined which cases had pairs of planets closer than 1.5 resonance widths to mutual first-order resonances (using equation~\ref{eq:reswidth} to calculate widths for low-eccentricity orbits consistent with our simulation initial conditions). 
We excluded these cases (which represent instances only $\sim5$ of the $\sim60$ observed planetary system architectures on which the simulations are based).
The stability of the remaining systems as a function of spectral fraction are shown in the bottom panels of Figure~\ref{fig:spectral-fractions}.
This removes the majority of the clear outliers; it is then clear that in systems not strongly influenced by MMRs, a spectral fraction above $\sim0.01-0.02$ in short simulations is correlated with a much lower chance of stability in long simulations. 
This supports the hypothesis that secular chaos is a significant driver of long term instability.
The location of the solar system's terrestrial planets near the secular stability boundary in the spectral fraction stability maps also supports this conclusion.

\begin{figure}
    %\centering
    \includegraphics[width=3.25in]{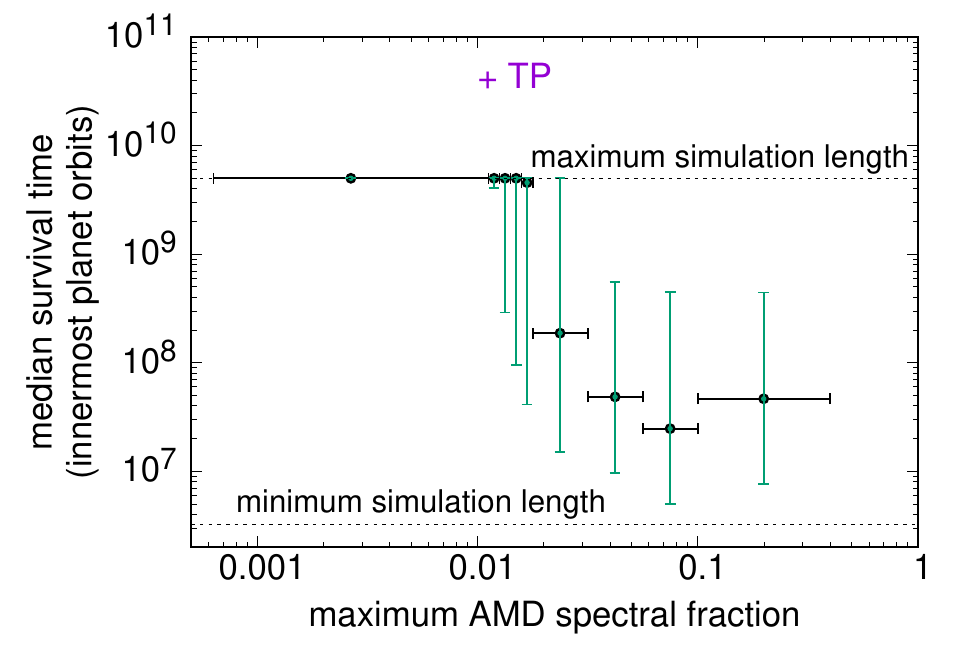}
    \caption{Median survival time (black points) for simulated planetary systems that do not have planets near/in 1st order mean motion resonances as a function of the maximum spectral fraction for the AMD of individual planets in the system. The horizontal black bars indicate the size of the spectral fraction bins (which is variable to ensure a sufficient sample of at least $\sim100$ simulations in each bin). The widest spectral fraction bin spans the full range of spectral fractions for which no system went unstable. The blue vertical lines show the range of survival times for the middle 68\% of the systems in each spectral fraction bin. Only simulations that were stable for $>3\times10^6$ orbits were included in the analysis. The spectral fraction for the solar system's terrestrial planets is indicated by `TP'; the y-axis value for the terrestrial planets is equivalent to $\sim10$ billion years, at which point \cite{Laskar:2009} estimate a $\sim1\%$ chance of Mercury obtaining a large eccentricity consistent with crossing Venus' orbit. 
}
    \label{f:sn-times}
\end{figure}

As a final step, we examine whether the stability timescales of the non-resonant planetary systems (from the bottom panels of Figure~\ref{fig:spectral-fractions}) are correlated with the AMD spectral fractions. 
The median survival time for a planetary system (in the long simulations) as a function of the AMD spectral fraction (from the short simulations) is shown in Figure~\ref{f:sn-times}. 
We find a rather sharp transition at a spectral fraction of 0.01:  below this value all of the cases are stable for the full $5\times10^9$ orbital periods of the long simulations; above this value, survival times generally decrease.

The strong correlations between long term dynamical stability and AMD spectral fraction support the hypothesis that secular chaos is an important driver of the evolution of planetary systems. The ``spectral fraction'' approach is a very promising tool for quickly diagnosing the likelihood of dynamical instabilities in a planetary system. 
The short simulations on which the spectral fraction calculations are based typically take only $\sim5-10$ minutes per system to complete on a desktop computer, compared to the hundreds of cpu hours required to integrate an individual system for billions of orbits.
 A spectral fraction parameter might be particularly valuable in attempts to predict planetary system stability using machine learning \citep[e.g.,][]{Tamayo:2016}.

%%%%%%%%%%%%%%%%%%%%%%%%%%%%%%%%%%%%%%%%%%%
%%%%%%%%%%%%%%%%%%%%%%%%%%%%%%%%%%%%%%%%%%%
\section{Summary and Conclusions}\label{s:summary}
%%%%%%%%%%%%%%%%%%%%%%%%%%%%%%%%%%%%%%%%%%%
%%%%%%%%%%%%%%%%%%%%%%%%%%%%%%%%%%%%%%%%%%%

We have performed a large suite of numerical simulations of planetary systems based on the observational sample of {\it Kepler} and {\it K2} transiting systems with four or more confirmed planets. We find that $\sim20\%$ of the cases show dynamical instabilities (leading to planet-planet collisions) within $5\times10^9$ orbital periods of the innermost planet. 
This result indicates that dynamical instabilities are not uncommon for plausible variations on the observed planetary architectures.
We find that these instabilities occur at a wide range of the planets' dynamical spacings, indicating that the source of instabilities in many of these simulations is not simply a result of planets being closely packed.
We find a wide range of instability timescales in our simulations. 
Correlation analysis of the stable versus unstable systems reveals only that the unstable systems typically have smaller period ratios of adjacent planet pairs (Section ~\ref{s:keplermultis}).

The bulk simulations motivated us to take a close look at what drives the instabilities in these systems. 
In our case-study of the Kepler-102 system (Section \ref{s:casestudy}), we find that mean motion resonances are unlikely to be the initial trigger of instability.
Despite period ratios that appear close to low-integer ratios,  for plausible values of planet masses and orbital eccentricities, the widths of mean motion resonances are too narrow for resonance overlap to occur and to directly influence dynamical stability (Section~\ref{ss:MMRs}). 
However, we do note that only modest eccentricity increases (up to $e\sim0.1$) are required for the 4:3 resonance between the innermost pair of the Kepler-102 system to affect their dynamics.
In Section \ref{ss:secular-dynamics}, we report evidence that the inward transfer of AMD via secular chaos is the most likely driver of eccentricity growth in the orbits of Kepler-102's inner planets in our simulations. 
 An estimate of the total system AMD required to raise the eccentricity of Kepler-102 b enough to induce instability provides an approximate lower limit of $\sim0.1M_\Earth$ on the mass of planet b  (Fig.~\ref{f:amd-example}). 
 This conclusion is supported by the numerical evidence of inward transfer of AMD in long term unstable versions of the Kepler-102 system (Fig.~\ref{f:amd-slope}). 
 Even for very low initial  eccentricities ($e\sim0.02$), both the AMD-stability estimate and the long-term simulations indicate that cases where Kepler-102 b's mass exceeds  $\sim0.1M_\Earth$ are more favorable for long-term stability. 
 This lower limit on planet b's mass corresponds to Earth-like bulk densities, and excludes the much lower densities allowed by statistical mass-radius relationships.

In Section~\ref{s:spectra}, we performed a frequency analysis of the AMD and the orbital elements for the planets in our simulated systems in short integrations (spanning a few secular cycles) and classified them according to a "spectral fraction" that quantifies whether a planet's secular evolution is dominated by a few well-defined frequencies (small spectral fraction) or has a noisy power spectrum (large spectral fraction).
We find that for planetary systems lacking first order mean motion resonances (which is the majority of the mutiplanet systems considered here), small spectral fractions (below $\sim0.01$) of the AMD calculated from integrations of a few million orbital periods are strongly associated with long term stability (found in much more cpu-intensive simulations spanning billions of orbits). 
We also find that the median stability timescale generally decreases with increasing AMD spectral number. 
This supports the hypothesis that secular chaos (which allows the transfer of AMD between planets via near-degenerate or resonant secular frequencies) is the dominant driver of instabilities in many multiplanet systems.
We conclude that the spectral fraction approach  also provides a promising tool to predict the longer-term stability of planetary systems by means of computationally inexpensive short term numerical simulations.

%%%%%%%%%%%%%%%%%%%%%%%%%%%%%%%%%%%%%%%%%%%%%%%%%%%%%%%%%%%%%%%%%%%%%%%%%%%%%%%%%%%%%%%%%%%%%%%%
\facilities{Exoplanet Archive, ADS}
\software{rebound,numpy}

%%%%%%%%%%%%%%%%%%%%%%%%%%%%%%%%%%%%%%%%%%%%%%%%%%%
{\it Acknowledgements:}
This work was supported by NASA (grant 80NSSC18K0397).
This research has made use of the NASA Exoplanet Archive, which is operated by the California Institute of Technology, under contract with the National Aeronautics and Space Administration under the Exoplanet Exploration Program.
An allocation of computer time from the UA Research Computing High Performance Computing (HPC) at the University of Arizona is gratefully acknowledged.
This paper includes data collected by the Kepler and K2 missions. Funding for the Kepler and K2 missions is provided by the NASA Science Mission directorate.

\appendix

\section{Mean Motion Resonance Width Expressions}\label{a:res}

For completeness, here we reproduce the analytical resonance width expressions from \citet{Murray:1999} along with values for the coefficients in the expressions.
For an internal $(p+q):p$ resonance with $q>1$ between a planet on a circular orbit and a massless test particle with eccentricity $e$, the resonance width is given by
\begin{equation}\label{eq:highorder}
    \frac{\Delta P}{P} = \left( 12 \mu \alpha |f_d| e^q \right)^{1/2},
\end{equation}
where the coefficient $f_d$ is a function of $\alpha$, $p$, and $q$  \citep{Murray:1999}. 
For external resonances, the expression is identical except $\alpha$ is omitted. 
For an internal first order $(p+1):p$ resonance between a planet on a circular orbit and a massless test particle with eccentricity $e$, the resonance width is given by
\begin{equation}\label{eq:1storder}
    \frac{\Delta P}{P} = \pm \left( 12 \mu \alpha |f_d| e \right)^{1/2}\left( 1+\frac{\mu \alpha |f_d|}{27p^2e^3} \right) - \frac{\mu \alpha |f_d|}{3pe},
\end{equation}
and again the expression for external resonances is given by omitting $\alpha$ \citet{Murray:1999}.
Table~\ref{t:coeff} lists the values of $\alpha$ and $f_d$ for a number of first and second order resonances. 
To calculate additional coefficients for resonances up to 4th order, we have compiled the expressions for $f_d$ given in Appendix B of \citet{Murray:1999} into a python notebook available here: \url{https://github.com/katvolk/analytical-resonance-widths}.

\begin{deluxetable}{|c|c|c|c|c|}
\tablecaption{Coefficients for resonance width calculations}
    \tablehead{$p+q$ & $p$ & $\alpha$ & $f_{d,i}$ & $f_{d,e}$} 
    \startdata
2 & 1 & 0.629961 & -1.190494 & -1.007837\\
3 & 2 & 0.763143 & -2.025223 & -1.824964\\
4 & 3 & 0.825482 & -2.840432 & -2.633396\\
5 & 4 & 0.861774 & -3.649618 & -3.439022\\
6 & 5 & 0.885549 & -4.456143 & -4.243361\\
3 & 1 & 0.480750 & 0.598757 & 0.014215\\
5 & 3 & 0.711379 & 3.273807 & 0.134711\\
7 & 5 & 0.799064 & 7.870501 & 0.372024\\
9 & 7 & 0.845740 & 14.386605 & 0.724300\\
    \enddata
    \tablecomments{The values $f_{d,i}$ and $f_{d,e}$ correspond to the values for internal and external resonances, respectively, calculated from the expressions in Appendix B of \cite{Murray:1999}.
    \label{t:coeff}}
\end{deluxetable}

\section{Kepler-102 TTVs}\label{a:ttvs}

\cite{Hadden:2014} examined transit timing variations in the Kepler-102 system for planets c, d, and e (planet b was not included in the dataset that formed the basis of their analysis). 
They fit a TTV of $16\pm5.8$ minutes to the transits of Kepler-102 c assuming that these TTVs are caused by the 3:2 near-resonance between planets c and d.
They also fit a TTV of $1.4\pm0.5$ to the transits of Kepler-102 e assuming the 3:2 near-resonance between planets d and e.
\citet{Hadden:2014} then used an analytical model for TTV amplitudes induced by these first-order MMRs to estimate the mass of planet d to be $2.6^{+1.3}_{-1.1} M_{\earth}$.

Here we use {\sc{rebound}} to simulate potential TTVs in the Kepler-102 system over a $\sim4$ year time-span to show two things: 1) the analytical TTV calculations do not appear to be accurate when a planetary system has chains of near resonances (i.e., when the TTVs involve more than just a single resonance between a single pair of planets) and 2) the 4:3 near-resonance between planets b and c would likely induces a larger TTV signal in planet c than the  3:2 near-resonance with planet d.

To show the first point, we simulated the Kepler-102 system in the absence of planet b (i.e., considering the same set of planets as considered in \citealt{Hadden:2014}) as follows. 
We randomly selected a mass for planet d from a uniform range within the \cite{Hadden:2014} mass estimate ($2.6^{+1.3}_{-1.1} M_{\earth}$); the same was done for planet e from within its RV mass range.  
We selected planet c's mass from the range $0.15-0.4 M_{\earth}$, which is a plausible mass range for terrestrial-planet densities given its radius and radius uncertainty.
We assigned planet f a mass of $1M_{\earth}$ with no variation because it does not strongly influence the TTVs of the other planets.
For simplicity, we assumed a co-planar system.
The mean anomaly for each planet was randomly selected from $0-2\pi$, with each planet's longitude of perihelion ($\varpi$) then set such that the planets would cross the x-axis of the simulation (which we use to define the line-of-sight for transit midpoints) at the observed relative transit times listed in the exoplanet archive (for the epoch JD 2454967.1). 
The planets' initial osculating eccentricities were randomly selected from 0-0.05; the \cite{Hadden:2014} analysis of the Kepler-102 system assumed that free eccentricities in the system were low, so our simulations should generally not have lower free eccentricities than they assumed and thus should not underestimate the TTVs.
We then performed $\sim200$ simulations spanning $\sim4.5$ years and recorded every x-axis crossing time for each planet (to within a one minute resolution).
A linear trend was fit and subtracted from these simulated transit mid-points to yield the TTVs for each simulation.

\begin{figure}[ht]
    \centering
    \includegraphics[width=4in]{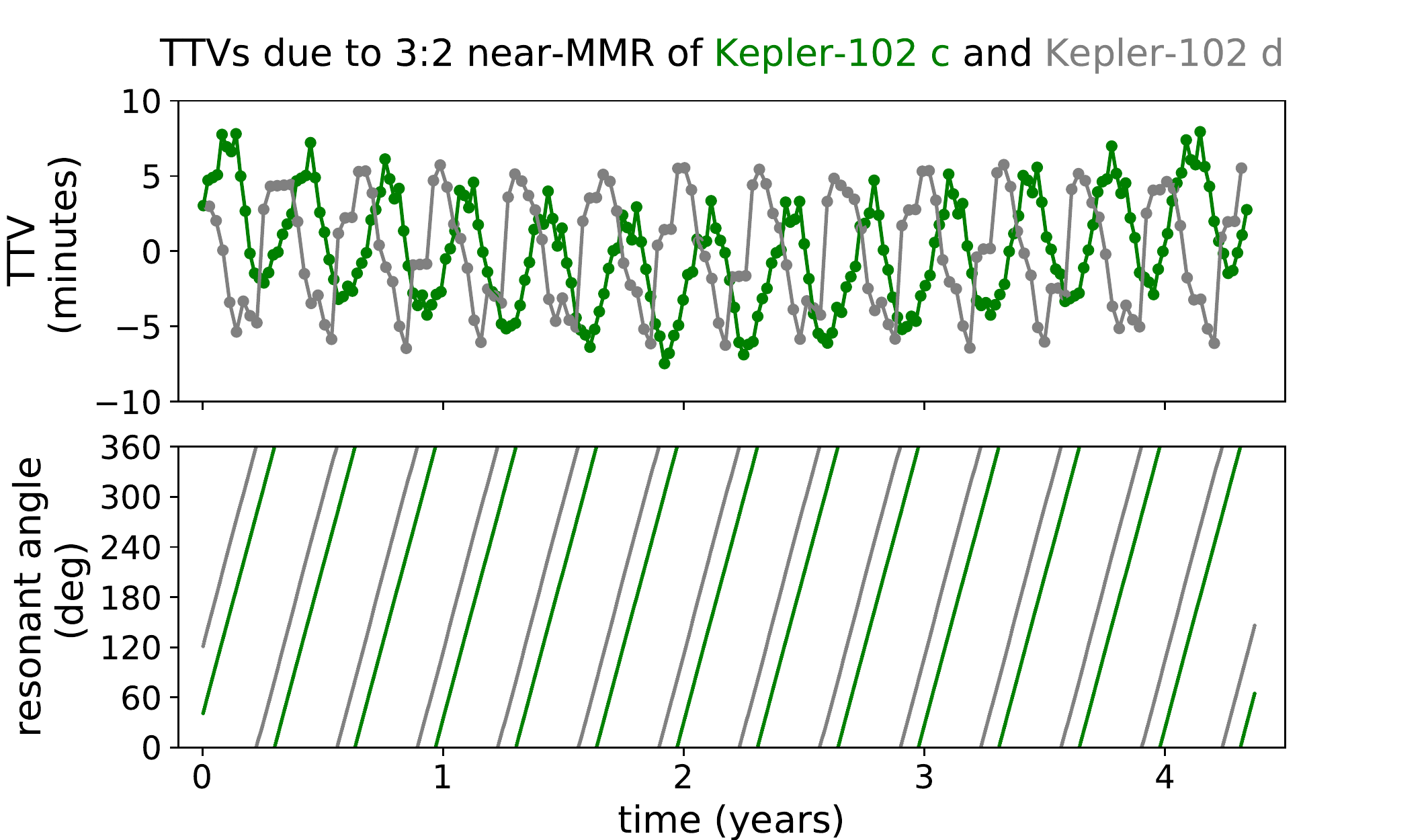}
    \caption{Simulated TTVs (top panel) for planets Kepler-102 c and Kepler-102 d in the absence of  Kepler-102 b. The dominant period in the simulated TTVs roughly matches the period for the circulation of the two possible eccentricity-type resonance angles (bottom panel) associated with a 3:2 mean motion resonance between planet c and d (gray: $3\lambda_d - 2\lambda_c - \varpi_d$; green: $3\lambda_d - 2\lambda_c - \varpi_c$).
    \label{f:cd32TTV}}
\end{figure}

Figures~\ref{f:cd32TTV} shows a representative TTV time series for planets c and d in these simulations along with the resonant angles for their near 3:2 MMR.
We find that the amplitude of Kepler-102 c's TTVs are typically only of order $\sim5$ minutes for \cite{Hadden:2014}'s nominal mass range, which was fit based on an apparent TTV signal of $\sim10-20$ minutes. We only find a few simulations where such a large TTV signal arises for planet c, and those tend to correspond to TTV signals for planet e (induce by the 3:2 near-resonance between d and e) that are too large compared to the $\sim1-2$ minute signal fit by \cite{Hadden:2014}.
We point this out to highlight that analytical TTV calculations that treat individual MMRs in isolation appear not to be sufficiently accurate in planetary systems such as Kepler-102 where there are chains of nearby resonances that contribute to the dynamical interactions. 
The analytical expressions accurately predict simulated TTV amplitudes for isolated pairs of planets near resonance (which we confirmed by simulating just planets c and d), but when we use these same expressions to predict planet c's TTV signal in a simulation with planets c, d, and e, the analytically derived amplitude rarely agrees with the simulations even to within a factor of two.

\begin{figure}[h]
    \centering
    \includegraphics[width=4in]{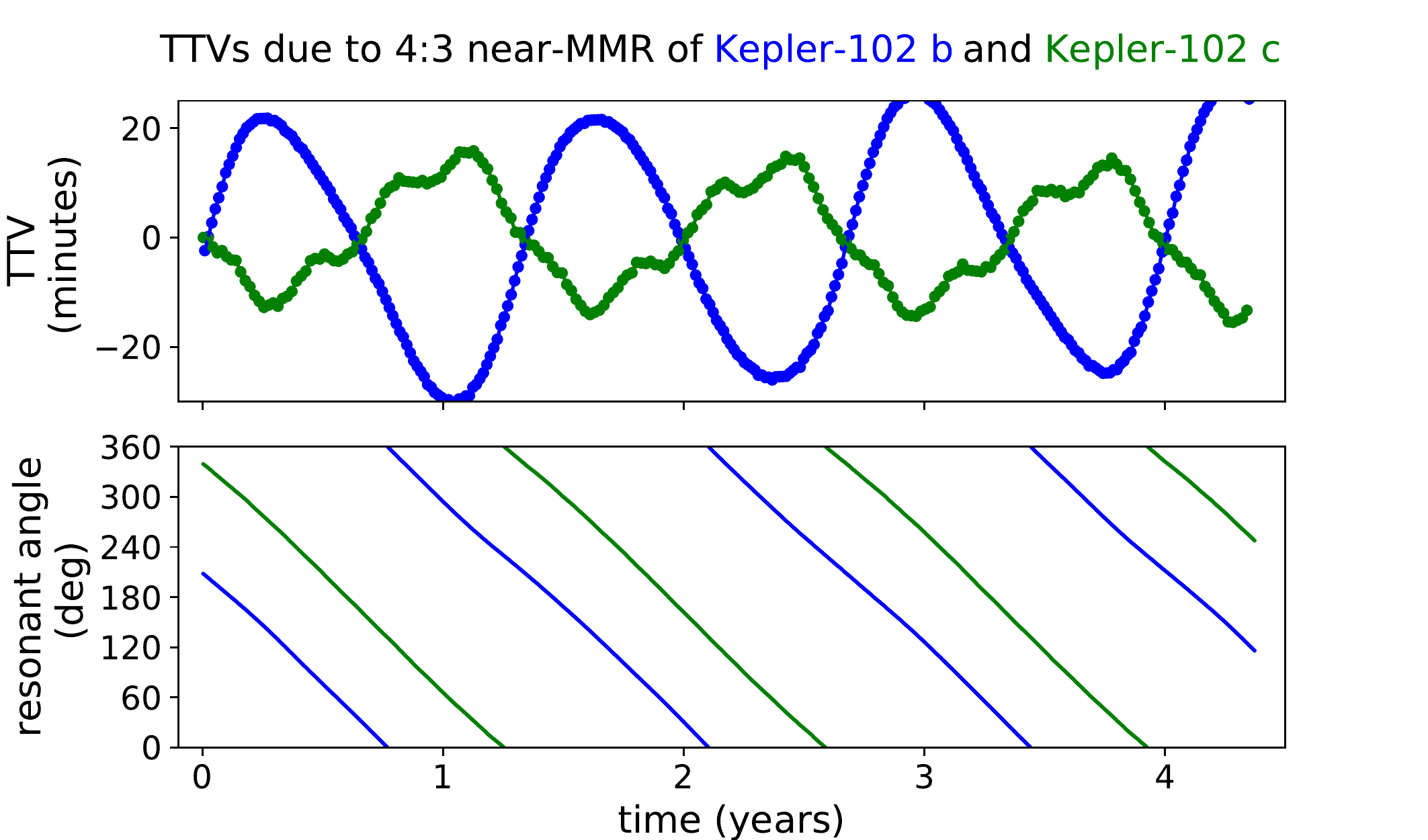}
    \caption{Simulated TTVs (top panel) for planets Kepler-102 b and Kepler-102 c. The dominant period in the simulated TTVs matches the period for the circulation of the two possible eccentricity-type resonance angles (bottom panel) associated with a 4:3 mean motion resonance between planet b and c (blue: $3\lambda_c - 2\lambda_b - \varpi_b$; green: $3\lambda_c - 2\lambda_b - \varpi_c$). There is a smaller additional TTV signal in the green curve for planet c; this signal is due to the 3:2 near-resonance between planets c and d (see Fig.~\ref{f:cd32TTV}).
    \label{f:bc43TTV}}
\end{figure}

Simulations also show that any potential TTVs exhibited by planet c are likely to be dominated or at least significantly influenced by the 4:3 near-resonance with planet b. 
We repeated the above described simulations with the addition of planet b (assuming a mass range of $0.075-0.225 M_{\earth}$).
In most of these simulations, the TTV signal of planet c shows a combination of periodic signals,: one at the period expected for the 4:3 near-resonance with b and one at the period expected for the 3:2 near-resonance with d; however the amplitude of the 4:3 near-resonance signal is typically 2-3 times larger than that for the 3:2 near-resonance (see figure~\ref{f:bc43TTV}).

\begin{figure}
    \centering
    \includegraphics[width=4in]{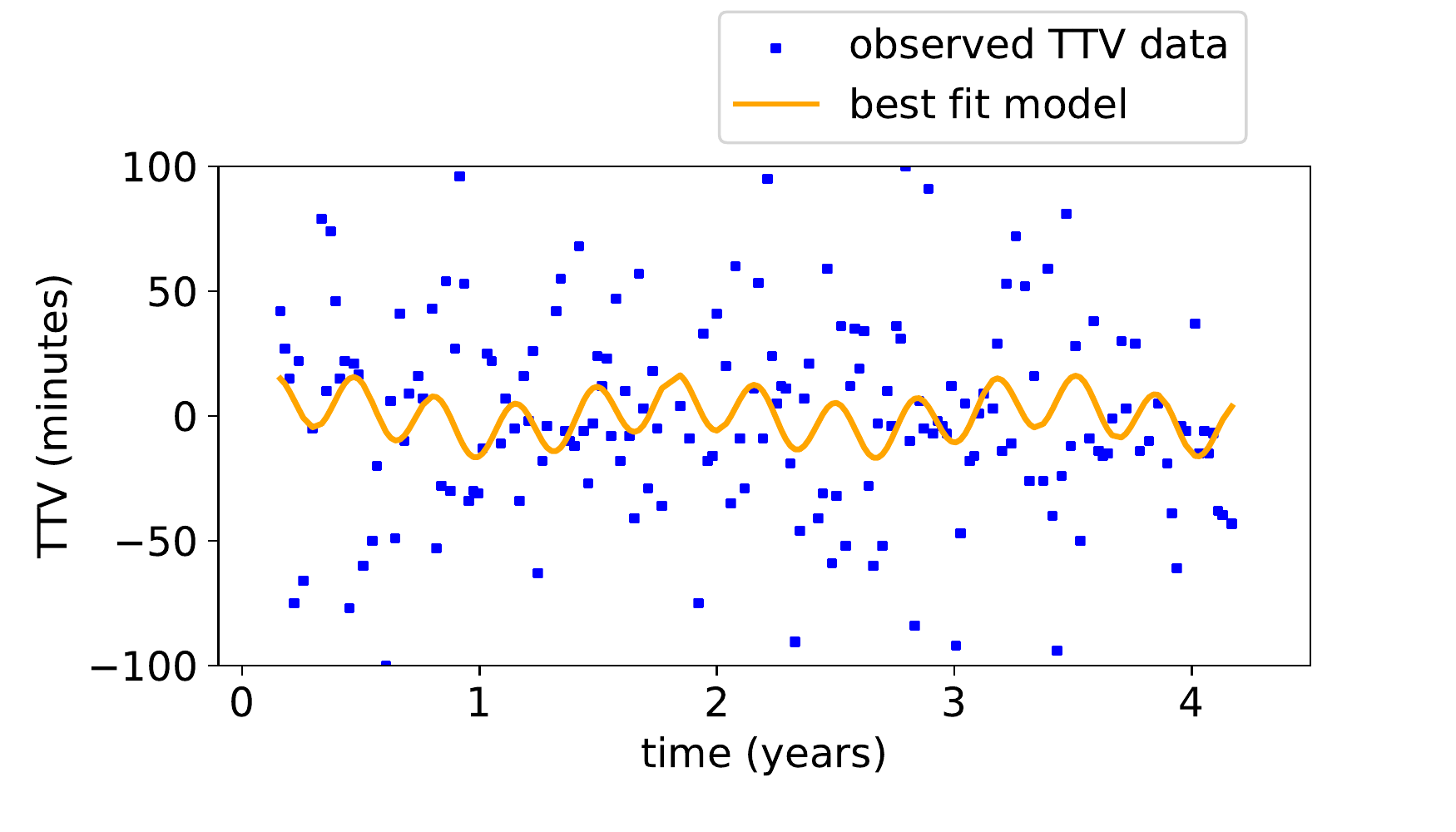}
    \caption{Observed transit timing variations for Kepler-102 c (blue squares; data from \citealt{Holczer:2016} with typical uncertainties of $\sim25$ minutes for each point) compared to a best-fit model for the TTVs (orange stars). The TTV model is a sum of two sinusoidal functions: one at the period for the 4:3 near-resonance between planets b and c, and one at the period of the 3:2 near-resonance between planets c and d. The fit is clearly not statistically significant.
    \label{f:obsttv}}
\end{figure}

Unfortunately, this discussion of TTVs for Kepler-102 is merely illustrative, because an analysis of TTVs from all available long-cadence {\it Kepler} data on Kepler-102 \citep{Holczer:2016} does not in fact find any significant periodic signals for the system.
We can confirm this finding by taking the time series of measured TTVs from  \cite{Holczer:2016} for the four outer planets in the Kepler-102 system and fitting sinusoidal TTVs at the expected periods for each of the near-resonances in the system; none of the fits are statistically significant (in every case the RMS of the residuals from the best-fit is nearly equal to the RMS of the data itself).
Figure~\ref{f:obsttv} shows the results for Planet c.

Thus, the mass for Kepler-102 d derived by \citet{Hadden:2014} should be disregarded in light of subsequent data. 
The TTV signals in the Kepler-102 system from the near-resonances do not seem significant \citep{Holczer:2016}.
The lack of significant signal might yield upper limits to some of the planet masses, but analytical methods will not suffice due to the multiple interacting resonances. 
We leave a numerical exploration of this as a potential future exercise.

\end{document}